\documentclass[twocolumn,twoside,slac_two]{revtex4}
\usepackage{graphicx}
\usepackage{fancyhdr}

\def\be{\begin{equation}}
\def\ee{\end{equation}}
\def\bea{\begin{eqnarray}}
\def\eea{\end{eqnarray}}

\begin{document}

\title{Recent Developments in Neutrino Phenomenology}

\author{A. Yu. Smirnov}
\affiliation{International Centre for Theoretical Physics,
Strada Costiera 11, 34100 Trieste, ITALY\\
Institute for Nuclear Research, RAS, 117312 Moscow, RUSSIA}

\begin{abstract}

The first phase of studies of the neutrino mass and mixing 
is essentially over. The outcome is the  discovery of non-zero neutrino 
mass and determination of the dominant structure 
of the lepton mixing matrix. 
In some sense this phase was very simple, and nature was very collaborative with us:  
Two main effects -  
the vacuum oscillations and the adiabatic conversion in matter (the MSW-effect) - 
provide complete  interpretation of the  experimental results. 
Furthermore,  with the present accuracy of measurements 
the $3\nu-$mixing analysis is essentially reduced to the $2\nu$ consideration. 
I will present a concise and comprehensive  description of this
first phase.  The topics  include: (i) the concept of neutrino mixing in vacuum and 
matter;  (ii) physics of the oscillations and adiabatic conversion; (iii) the experimental  
evidences of the flavor transformations  and determination of the oscillation parameters.    
Some implications of the obtained results are discussed. 
Comments are given on the next phase of the field that will be much more involved.   
 
\end{abstract}

\maketitle

\thispagestyle{fancy}

\section{Introduction}

Recent major progress in neutrino phenomenology,  
and particle physics in general, 
was related to  studies of the neutrino mass and mixing. 
The first phase of these studies is essentially over, 
with the  main results being  

\begin{itemize}

\item
discovery of non-zero neutrino mass;  

\item 
determination of the dominant structure of the lepton mixing:
discovery of  two large mixing angles;

\item 
establishing strong difference of the quark and lepton 
mass spectra and mixing patterns. 

\end{itemize}
 
Physics of this first phase is rather simple. 
The two main effects  -  the vacuum oscillations 
\cite{pontosc,mns,pontprob}
and  the adiabatic conversion in matter (the MSW-effect) \cite{w1,ms1}.
are enough for complete interpretation of 
the experimental results.  
(Oscillations in matter appear as sub-leading statistically insignificant yet 
effect for the atmospheric and solar neutrino oscillations in the Earth.)
Furthermore, at the present level of experimental accuracy  
the three neutrino analysis is essentially 
reduced  to two neutrino analyzes,     
and  degeneracy of parameters is practically absent. 
Nature was very ``collaborative'' with us, realizing  the easiest 
possibilities and  disentangling an interplay of various phenomena.

In a sense, we have now a ``standard model of neutrinos''
that can be formulated in the following way:

1).  there are only three types of  light neutrinos;

2).  their interactions are described by the Standard electroweak 
theory;  

3). masses and mixing are generated in vacuum;   
they originate from  some high energy (short range) physics 
at the electroweak or/and higher scales.  


Now the goal is to test these statements and  
to search for new physics beyond
this  ``standard model''.
Confirmation of the LSND result by MiniBooNE would be 
discovery of such a new physics. \\

The next phase of studies  will be associated 
to new generation of neutrino experiments,
which will start in 2008 - 2010. 
The main objectives of this new phase include 
determination the absolute scale of neutrino mass  
and sub-dominant structures of mixing: namely, 
1-3 mixing, deviation of the 2-3 mixing from maximal value,
the  CP-violation phase(s). 
The objectives  include also identification of neutrino mass hierarchy 
and precision measurements of already known parameters.

The next phase will be much more involved:
New phenomena may show up at the sub-leading level. 
More complicated formalisms for their interpretation
are required. 
Complete three-neutrino context of study will be the must. 
Severe problem of degeneracy of parameters appears.\\

In these lectures\footnote{The text presented here is partially based on 
lectures given at the  Les Houches Summer School on 
Theoretical Physics: Session 84: Particle 
Physics Beyond the Standard Model, Les Houches, France, 1-26 Aug 2005, as well as on  materials 
prepared for the TASI-06 school ``Exploring New Frontiers Using Colliders and Neutrinos'', June 4 - 
30, 2006, Boulder Colorado.} 
I will present a concise description of the first phase 
of studies of neutrino masses and mixing. 
I will start by a detailed  discussion of the concept of neutrino mixing 
in vacuum and matter. 
In the second part, 
the the main effects involved: the vacuum oscillations, oscillations in matter and 
the adiabatic conversion, are described  
and physics derivation of all relevant formulas are given. 
In the third part I will present the experimental results  
and  existing evidences of neutrino oscillations. 
For each experiment a simple analysis is described that allows 
one to evaluate the  neutrino parameters without 
sophisticated global fit. 
This consideration is aimed at convincing  that indeed,  we see the oscillations and 
and our interpretation of results 
in terms of the vacuum masses and mixing is correct.




\section{Flavors,  masses and  mixing}

\subsection{Flavor mixing}

The {\it flavor} neutrinos,  $\nu_f \equiv (\nu_e, \nu_{\mu}, \nu_{\tau})$ 
are defined as the neutrinos that  correspond  to certain charge leptons:  
$e$, $\mu$ and $\tau$. The correspondence is  established by  
the weak interactions: $\nu_{l}$ and $l$ ($l = e, \mu, \tau$)  
form  the charged  currents or doublets of the $SU_2$ symmetry group. 
Neutrinos,  $\nu_1$, $\nu_2$, and $\nu_3$,  
with definite masses $m_1$, $m_2$, $m_3$ 
are the eigenstates of mass matrix as well as the eigenstates of 
the total Hamiltonian in vacuum. 

The {\it vacuum mixing}  means that the flavor states 
do not coincide with the mass eigenstates.  
The flavor states are  combinations of the mass eigenstates: 
\be 
\nu_{l} = U_{l i} \nu_{i}, ~~~ l = e, \mu, \tau, ~~~i = 1, 2, 3,  
\label{mixing}
\ee
where the mixing parameters  $U_{l i}$ form the  PMNS mixing matrix 
$U_{PMNS}$ \cite{pontosc,mns}.  
The mixing matrix can be conveniently parameterized as 
\be 
U_{PMNS} = V_{23}(\theta_{23}) I_{- \delta} V_{13}(\theta_{13}) 
I_{\delta} V_{12}(\theta_{12}),  
\ee
where $V_{ij}$ is  the rotation matrix in the $ij$-plane, 
$\theta_{ij}$ is the corresponding angle and 
$I_{\delta} \equiv diag(1, 1, e^{i\delta})$
is the matrix of CP violating phase.

\subsection{Two aspects of mixing.}

A number of  conceptual points can be clarified using just $2\nu-$ mixing. 
Also, at the present level of accuracy of measurements 
the $2\nu-$ dynamics is enough to describe the data. 
For two neutrino mixing, {\it  e.g.}  $\nu_e - \nu_a$, 
we have    
\begin{eqnarray}
\nu_e = \cos\theta \nu_1 + \sin \theta \nu_2, ~~
\nu_a = \cos\theta \nu_2 - \sin \theta \nu_1, 
\label{2nu}
\end{eqnarray}
where $\nu_a$ is the non-electron neutrino state,  
and $\theta$ is the vacuum mixing angle.

There are two important aspects of  mixing. 
The first aspect: according to 
(\ref{2nu}) the flavor neutrino states are
combinations of the mass eigenstates. 
Propagation of 
$\nu_e$ ($\nu_a$) is described by a system of two wave packets 
which correspond to  $\nu_1$ and $\nu_2$. 
\begin{figure}[htb]
\centering
\includegraphics[width=80mm,angle=0]{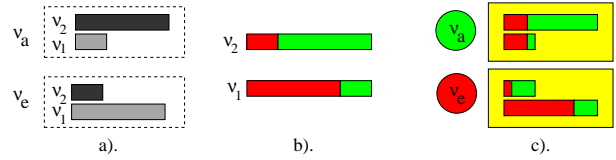}
\caption{a). Representation of the flavor neutrino states   
as the combinations of the mass eigenstates. 
The length of the box gives the admixture of (or probability to 
find) corresponding mass state in a given flavor state. 
(The sum of the lengths of the boxes is normalized to 1. 
b). Flavor composition of the mass 
eigenstates. The electron flavor is shown by red 
(dark) and the non-electron flavor by green  (grey). 
The sizes of the red and green parts give the 
probability to find  the electron and non-electron neutrino in a given 
mass state.
c). Portraits of the electron   and non-electron neutrinos: shown are 
representations 
of the electron and non-electron  neutrino states  as combinations of the 
eigenstates  for which, in turn, we show  the flavor composition.}
\label{mix}
\end{figure}
In fig.~\ref{mix}a). we show representation 
of $\nu_e$ and $\nu_a$ as the combination of mass states. 
The lengths of the boxes,  $\cos^2\theta$ and $\sin^2\theta$,  give 
the {\it admixtures } of $\nu_1$ and $\nu_2$ in $\nu_e$ and $\nu_a$.  
The key point is that the flavor states are {\it coherent} mixtures 
(combinations) of the 
mass eigenstates. The {\it relative phase} or phase difference 
of $\nu_1$ and $\nu_2$ in $\nu_e$ as well as  $\nu_a$ is fixed:   
according to  (\ref{2nu}) it is zero in $\nu_e$ and $\pi$  in $\nu_a$.  
Consequently,  there are 
certain {\it interference}  effects between $\nu_1$ and $\nu_2$ which 
depend on the relative 
phase.

The second aspect: the relations (\ref{2nu}) can be inverted: 
\be
\nu_{1} = \cos\theta~\nu_e - \sin \theta~\nu_{a}, ~~
\nu_{2} = \cos\theta~\nu_{a} + \sin \theta~\nu_{e}.
\label{2nuin}
\ee
In this form they determine the {\it flavor composition} 
of the mass states (eigenstates of the Hamiltonian), 
or shortly, the flavors of eigenstates. According to  (\ref{2nuin}) 
the probability  to find the electron flavor in $\nu_{1}$ is given by 
$\cos^2\theta$, whereas 
the probability  that $\nu_{1}$ appears 
as $\nu_a$ equals $\sin^2\theta$. This flavor decomposition is 
shown in fig.~\ref{mix}b). by colors (different shadowing).

Inserting the flavor decomposition of mass states 
in the representation of the flavors states,  
we get  the  ``portraits" of the electron and non-electron neutrinos fig.~\ref{mix}c). 
According to this figure,  $\nu_e$ is a system of  
two mass eigenstates that, in turn,   have a composite flavor. 
On the first sight the portrait has a paradoxical feature: there is the 
non-electron (muon and tau) flavor 
in the electron neutrino!  The paradox has the following resolution: 
in the $\nu_e$- state   
the $\nu_a$-components of  $\nu_{1}$ and $\nu_{2}$ are equal and have opposite 
phases. 
Therefore they cancel each other and 
the electron neutrino has pure electron flavor as it should be. 
The key point is  interference: 
the interference of the non-electron parts is destructive in $\nu_e$. 
The electron neutrino has a ``latent"  non-electron component 
which can not be seen due to particular phase arrangement. However,  during propagation 
the phase difference changes 
and the cancellation disappears. This leads to an appearance of the 
non-electron component 
in  propagating neutrino state which was originally 
produced as the electron neutrino. This  is the mechanism 
of neutrino oscillations. Similar consideration holds for the $\nu_a$  state.

\subsection{Who mixes neutrinos?} 

How mixed neutrino states (that is, the coherent mixtures on the mass 
eigenstates) are created? Why neutrinos and not charged leptons? 
In fact, these are non-trivial questions.  
Creation (preparation - in quantum mechanics terms) of the 
mixed neutrino states is a result of interplay of the charged current 
weak interactions and kinematic features of specific reactions. 
Differences of masses of the charged leptons play crucial role. 

Let us consider three neutrino species separately. 

1).  Electron neutrinos: The combination of  mass eigenstates,  
which we call the electron  neutrino,  is produced,  
{\it e.g.}, in the beta decay (together with electron). 
The reason is the energy conservation:  
no other combination can be produced  because the energy release 
is  about few MeV,  so that neither muon nor tau lepton can appear. 


2). Muon neutrino. Almost pure $\nu_{\mu}$ state is produced together 
with muons in the charged  pion decay: $\pi^+ \rightarrow \mu^+ \nu_{\mu}$. 
Here the reason is ``chirality suppression'' - essentially 
the angular momentum conservation and  V-A  character of the 
charged current weak interactions.  
The amplitude is proportional to the mass of the charged lepton 
squared.  Therefore the channel with the electron neutrino: 
$\pi^+ \rightarrow e^+ \nu_e$ is suppressed as $\propto m_e^2/m_{\mu}^2$.  
Also coherence between $\nu_{\mu}$ and small admixture of $\nu_e$ is lost 
almost immediately due to difference of kinematics.

3).  Tau neutrino. Enriched $\nu_{\tau}$ - flux  can be 
obtained in the beam-dump experiments at high energies: 
In the thick target all light mesons 
($\pi$, $K$  which are sources  of usual neutrinos)   
are absorbed  before decay, 
and only heavy short living particles, like $D$ mesons,  
have enough time to decay. 
The $D$ mesons have also modes of decay with emission of 
$\nu_e$ and $\nu_{\mu}$ that  are  chirality-suppressed 
in comparison with $D \rightarrow \tau \nu_{\tau}$.   
Furthermore,  coherence of $\nu_e$ and $\nu_{\mu}$ with 
$\nu_{\tau}$ is lost due to strongly different energies and momenta. 

What about the neutral currents? 
Which  neutrino state is produced in the $Z^0-$decay in the presence of mixing?
$Z^0-$interactions are flavor blind and all the neutrino
flavors are produced  with the same amplitude (rate). The only
characteristic that distinguishes neutrinos is the mass.
So, the state produced in the $Z^0$-decay can
be written as 
\be
| f \rangle = \frac{1}{\sqrt{3}}
\left[
|\bar{\nu}_1 \nu_1 \rangle  +
|\bar{\nu}_2 \nu_2 \rangle  +
|\bar{\nu}_3 \nu_3 \rangle
\right]
\ee
(which is also equivalent to the sum of pairs of the flavor states) \cite{gt}. 
It is straightforward to show that  the
decay rate $|Z^0 \rangle \rightarrow |f \rangle $  is given by
\be
|\langle f | H | Z^0 \rangle|^2 =
3|\langle  \bar{\nu}_1 \nu_1 |H| Z^0 \rangle|^2, 
\ee
that  coincides with what one obtains in the case of three
independent decay channels.

Do neutrinos from $Z^0$-decay oscillate?
One can show that oscillations can be 
observed in the two-detector experiments
when both neutrinos from the decay are detected \cite{gt}.
If a flavor of  neutrino (antineutrino)  is fixed,
then a flavor of the accompanying  antineutrino (neutrino) will  oscillate 
with distance and energy.


\section{Physics effects}

\subsection{To determination of oscillation parameters}

In the Table \ref{tab1} we show parameters to be determined, 
sources of information for their determination 
and the main physical effects involved.  
In the first approximation, when 1-3 mixing is neglected, 
the three neutrino problem splits into two neutrino problems 
and parameters of the 1-2 and 2-3 sectors can be determined independently. 

\begin{table}[b]
\caption{Parameters and effects.}
\label{tab1}
\tabcolsep=0.1cm
\begin{tabular*}{\columnwidth}{@{\extracolsep{\fill}}@{}lll@{}}
\hline
Parameters   &  Source of information  &  Main physics effects \\ 
\hline
 $\Delta m_{12}^2$, $\theta_{12}$  &  Solar neutrinos &   Adiabatic conversion\\ 
 \,                        &  \,                 & and averaged vacuum\\ 
\,                         &  \,                 &    oscillations \\
\,                         &  KamLAND         & Non-averaged vacuum\\ 
\,                         &  \,                &     oscillations \\
\hline
$\Delta m_{23}^2$, $\theta_{23}$ & Atmospheric neutrinos & Vacuum oscillations \\
\,                               & K2K                 &  Vacuum oscillations\\
\hline
$\theta_{13}$   & CHOOZ  &   Vacuum oscillations \\
\,   &   Atmospheric neutrinos & Oscillations in matter \\
\hline
\end{tabular*}
\end{table}

Essentially two effects are relevant for interpretation 
of the present data in the lowest approximation: 

\begin{itemize} 
\item
vacuum oscillations (both averaged and non-averaged)\cite{pontosc,mns,pontprob}; 

\item 
adiabatic conversion in medium \cite{w1,ms1}. 

\end{itemize}

{\it A priory} another effect - oscillations in matter -  
should also be used in the analysis. It is relevant for 
the solar and atmospheric neutrinos propagating 
in the matter of the Earth.  It happens however,  that 
for various reasons the effect is small - at 
$(1 - 2)\sigma$ level and can be neglected in the first approximation. 

In the case of solar neutrinos, for the preferable values of 
oscillation parameters of the LMA solution (see below) 
this effect is indeed small.  Furthermore,  due to 
the attenuation (see below)  the Earth-core effect is small and 
one can consider oscillations as ones in constant density. 

In the case of atmospheric neutrinos 
the  $\nu_e -$ and $\nu_\mu -$  transition probabilities 
driven by 1-2 mixing and mass splitting are not small 
(of the order one in the sub-GeV range). However,  due to an  
accidental coincidence (the fact that the ratio of the muon-to-electron 
neutrino fluxes equals 2)  
the effect cancels for maximal 2-3 mixing (see below).

Notice also that the $2\nu-$ mixing analyzes are enough. However,  
in the next order, when sub-leading effects are included, 
the problem becomes much more difficult and degeneracy 
of parameters appear. We will comment on this later.

\subsection{Neutrino oscillation in vacuum}

In vacuum, the neutrino mass states are the eigenstates of the Hamiltonian. 
Therefore dynamics of propagation has the following features: 

\begin{itemize}

\item
Admixtures of the eigenstates (mass states) in a given neutrino state do not change. 
In other words, there is no  $\nu_1 \leftrightarrow \nu_2$ transitions.  
$\nu_1$ and  $\nu_2$ propagate independently.    
The admixtures are determined by mixing in a production point 
(by $\theta$,  if pure  flavor state is produced). 

\item 
Flavors of the eigenstates  do not change. They are also determined by $\theta$. 
Therefore the picture of neutrino state (fig. \ref{mix} c) does not change during 
propagation. 

\item 
Relative phase (phase difference) of the eigenstates monotonously increases. 

\end{itemize}

The phase is the only operating degree of freedom 
and we will consider it in details.  

{\it Phase difference.} 
Due to  difference of masses, the states  $\nu_1$ and  $\nu_2$ have different 
phase velocities 
$v_{phase} = E_i/p_i  \approx 1 + m_i^2/2E^2$
(for ultrarelativistic neutrinos), so that
\be
\Delta v_{phase} \approx \frac{\Delta m^2}{ 2E^2},~~~ \Delta m^2 \equiv m_2^2 - m_1^2. 
\ee
The phase difference changes as
\be
\phi = \Delta v_{phase} E t.
\label{phdif}
\ee

Explicitly, in the plane-wave approximation we have 
the phases  of two mass states  $\phi_i = E_i t - p_i x$. 
Apparently,  the phase difference 
which determines the interference effect
one should be taken in the same space-time point:
\be
\phi  \equiv \phi_1 - \phi_2 =    \Delta E t -  \Delta p x.
\label{phased}
\ee
Since $p = \sqrt{E^2 - m^2}$, we have
\be
\Delta p = \frac{dp}{ dE} \Delta E + \frac{dp}{ dm^2} \Delta m^2 =
\frac{1}{v_g}\Delta E  - \frac{\Delta m^2}{2p}, 
\label{deltap}
\ee
where $v_g = dE/dp$ is the group velocity.  Plugging (\ref{deltap})
into (\ref{phased}) we obtain 
\be
\phi  =   \Delta E \left( t - \frac{x}{v_g}\right) +
 \frac{\Delta m^2}{2p} x.
\label{phased2}
\ee
Depending on physical conditions either $\Delta E \approx 0$
or/and  $(t - x/v_g)$ is small which imposes the bound on size of the wave packet.
As a consequence, the first term is small and
we reproduce the result (\ref{phdif}).
For stationary source one should take
$\Delta E = 0$.

In general, depending on conditions of production and detection
both quantities $\Delta E$  and $\Delta p$ are non-zero.
There is  always certain time interval
in the problem, $\Delta t$,  that determines 
(according to the uncertainty principle) 
the energy interval $\Delta E$. {\it E.g.} 
in the case of solar neutrinos we know a time interval 
(determined  by the time resolution of a detector) when 
a given neutrino is detected.  
Furthermore,  neutrino production processes 
have certain life-times, or coherence times. 
There are arguments that one should take
the center of the wave packet where   $t = x/v_g$, 
or average over the wave packet length that leads to 
vanishing the first term in (\ref{phased2}). 
In both cases  one  obtains standard expression
for the phase. Apparently, the oscillation effect should disappear
in the limit $\Delta m^2 = 0$.


Notice that oscillations  are the effect
in the configuration space. 
The process is  described by  interference of
the wave functions that correspond to the mass
eigenstates, $\psi_1(x,t)$  and $\psi_2(x,t)$.
Formally, we can perform  the Fourier expansion of these wave functions
considering the  interference  in the momentum representation.
So,  formally we can always take the same momenta
doing then appropriate integration.\\

Increase of the phase leads to the oscillations. Indeed, the change of  phase  
modifies  the interference: in particular, cancellation of the non-electron   
parts in the state produced as $\nu_e$ disappears  and the non-electron component 
becomes observable.  
The process is periodic: when $\phi = \pi$,   the interference of 
non-electron parts is constructive 
and at this point the probability to find $\nu_a$ is maximal. 
Later, when $\phi = 2\pi$, 
the system returns to its original state: $\nu(t) = \nu_e$. 
The oscillation length is the distance at which this return occurs: 
\be
l_{\nu} = \frac{2\pi}{\Delta v_{phase} E} = \frac{4 \pi E}{\Delta m^2}. 
\ee
The depth of oscillations,  $A_P$,  is determined by the mixing angle. 
It is given by maximal probability  to observe the  ``wrong" flavor $\nu_a$. From 
the fig. 1c. 
one finds immediately (summing   up  the parts with the non-electron flavor in 
the amplitude) 
\be
A_P =  (2 \sin\theta \cos\theta)^2 = \sin^2 2\theta. 
\ee
Putting things together we obtain expression for the transition probability
\be
P = A_P \left(1 - \cos \frac{2\pi L}{l_{\nu}} \right) = \sin^2 2\theta 
\sin^2 \frac{\Delta m^2 L}{4E}. 
\label{oscprob}
\ee

The oscillations are the effect of the phase increase 
which changes the interference pattern.  
The depth of oscillations is the measure of mixing.

\subsection{Paradoxes of neutrino oscillations}

A number of issues in theory of neutrino oscillations is still 
under discussion.  Here I add several comments. 
Naive plane-wave description
reproduces correct result  
since it catches the main feature of the effect:
phase difference change.
Clearly it can not explain whole the picture because 
the oscillations are a {\it finite} space-time effect. 

Field theory approach 
provides with a consistent description.
Oscillation experiment  includes
neutrino production in the source,
propagation between the source and detector,
detection. 
In this approach production, propagation and detection of neutrinos are  
considered as a unique process in which  $\nu_1$ and
$\nu_2$ are virtual particles propagating between
the production, $x_P$,  and detection, $x_D$,  points.
Propagation of $\nu_i$ ($i = 1, 2$) is described by propagators
$S_i(x_D - x_P)$. 
Notice that here there is a substantial difference from our standard
calculations of the probabilities and cross-sections when
we consider the {\it  asymptotic states} and perform
integration over the  infinite space-time. The later leads to appearance
of the delta-functions that express conservation of the energy and
momentum. In the case of oscillations integration should be performed
over finite production and detection regions (integration over
$x_P$ and $x_D$). Also one should take into account finite
accuracy of measurements of the energy and momenta
of external particles.

From this point of view in usual consideration we perform 
{\it truncation} of whole process:
For $|x_P - x_D| \gg 1/\Delta p$ neutrinos can be considered
as real (on-shell) particles with negligible corrections due to
virtuality. 
Whole the process  can be truncated in three parts:
1).  production;  
2). propagation, as propagation of wave packets;
3). detection.
Neutrino masses are neglected in the production and 
detection processes. 
In this picture,  the oscillations are considered 
as the  effect of
propagation with certain initial and final conditions that
reflect process of production and detection. (Their effects develop over 
much larger space-time intervals.)
Correct boundary (initial and  final) conditions should be imposed. 
Essentially these conditions determine  the length and shape
of the wave packets. 

Let us stress again that oscillations are  the finite space and
finite time phenomenon: all the phases of the processes,
production, propagation and detection occur
(and should be considered) in the finite time intervals and
finite regions of space.


\subsection{Evolution equation} 

In vacuum the mass states are the eigenstates of Hamiltonian.  
So, their propagation is described by independent equations 
\be
i \frac{d\nu_i}{dt} = E_i \nu_i \approx \left(p_i + \frac{|m_i|^2}{2p_i}\right) \nu_i, 
\ee 
where we have 
taken ultrarelativistic limit and omitted the spin variables 
that  are irrelevant for the flavor oscillations. In the matrix 
form for three neutrinos $\nu \equiv (\nu_1, \nu_2, \nu_3)^T$, 
we can write
\be
i\frac{d \nu}{dt} \approx \left(p I + \frac{|M_{diag}|^2}{2E}\right)\nu, 
\ee
where $M_{diag}^2 = diag(m_1^2, m_2^2, m^2_3)$. 
Using the relation $\nu = U_{PMNS}^{\dagger}\nu_f$ (\ref{mixing}),  
we obtain 
the equation for the flavor states: 
\be
i\frac{d \nu_f }{dt} \approx  \frac{ M^2}{2E} \nu_f, 
\label{flevol}
\ee
where $M^2 \equiv U_{PMNS} |M_{diag}|^2 U_{PMNS}^{\dagger}$ 
is the mass matrix squared in the flavor basis. 
In (\ref{flevol}) we have omitted the  term 
proportional to the unit matrix which does not produce any phase difference 
and can be absorbed in the renormalization of the  neutrino wave functions. 
So, the Hamiltonian of the neutrino system in vacuum is 
\be
H_0 = \frac{|M|^2}{2E}.
\label{Hvacuum}
\ee
In the $2\nu$ mixing case we have explicitly: 
\be
H_0 = \frac{\Delta m^2}{4E}
\left(\begin{array}{cc}
-\cos 2 \theta  &  \sin 2 \theta \\
\sin 2 \theta &  \cos 2 \theta  
\end{array}
\right). 
\label{Hvacuum2}
\ee
Solution of the equation (\ref{flevol}) with this Hamiltonian  
leads to the standard oscillation formula (\ref{oscprob}).


\subsection{Matter effect}

{\it Refraction}. In matter,   neutrino propagation  is affected by interactions. 
At low  energies the  {\it  elastic forward scattering} is relevant only 
(inelastic interactions can be neglected) \cite{w1}.   
It can be described  by the potentials $V_e$, $V_a$. In usual medium 
difference of the  potentials for $\nu_e$ and $\nu_a$ 
is due to  the charged current scattering 
of  $\nu_e$ on electrons  ($\nu_e e \rightarrow \nu_e e$) \cite{w1}: 
\be
V = V_e - V_a  = \sqrt{2} G_F n_e~,
\ee
where $G_F$ is the Fermi coupling constant 
and $n_e$ is the number density of electrons. 
The result follows straightforwardly from calculation of the matrix element 
$V = \langle \Psi | H_{CC}| \Psi \rangle$, where $\Psi$  
is the state of medium and neutrino. 
Equivalently, one can describe the effect of medium in terms of the refraction index: 
$n_{ref} - 1 = {V}/{p}$.  

The difference of the potentials leads to an appearance of  
additional phase difference in the neutrino system: 
$\phi_{matter} \equiv (V_e - V_a) t$. 
The difference of potentials (or refraction indexes) determines 
the {\it refraction length}: 
\be 
l_0 \equiv \frac{2\pi}{V_e - V_a} = \frac{\sqrt{2}\pi} {G_F n_e} . 
\ee
$l_0$  is the distance over   which an additional ``matter" phase  equals $2\pi$. \\

In the  presence of matter the Hamiltonian of system changes: 
\be
H_0  \rightarrow H = H_0 + V ,
\ee
where $H_0$ is the Hamiltonian in vacuum.  
Using (\ref{Hvacuum}) we obtain (for $2\nu$ mixing)
\be
H  =  \frac{|M|^2}{2E} + V, ~~~~ V = diag(V, 0). 
\label{Hmatt}
\ee

The evolution equation for the flavor states in matter 
then becomes 
\be
i\frac{d\nu_f}{dt} =   \left[ \frac{\Delta m^2}{4E}
\left(\begin{array}{cc}
- \cos 2 \theta  &  \sin 2 \theta \\
\sin 2 \theta &  \cos 2 \theta  
\end{array}
\right) + V \right]\nu_f.  
\label{eqmatt2}
\ee

The eigenstates and the eigenvalues 
change: 
\begin{eqnarray}
\nu_1,~~ \nu_2 ~~~\rightarrow ~~~\nu_{1m}, ~~\nu_{2m},\\
\frac{m_1^2}{2E},~~ \frac{m_2^2}{2E}~~ \rightarrow ~~ H_{1m},~~ H_{2m} . 
\end{eqnarray}
The mixing in matter 
is determined with respect to the eigenstates 
of the Hamiltonian in matter,  $\nu_{1m}$ and $\nu_{2m}$. 
Similarly to (\ref{2nu}) the mixing angle in matter, $\theta_m$,  
gives  the relation between the eigenstates in matter and  the flavor states: 
\bea
\nu_e = \cos\theta_m \nu_{1m} + \sin \theta_m \nu_{2m}, 
\nonumber \\
\nu_a = \cos\theta_m \nu_{2m} - \sin \theta_m \nu_{1m}.
\label{2num}
\eea
The angle $\theta_m$  in matter is obtained by diagonalization 
of the Hamiltonian in matter (\ref{Hmatt}):
\be
\sin^2 2\theta_m = 
\frac{\sin^2 2\theta}{(\cos 2\theta - 2VE/\Delta m^2 )^2 +  \sin^2 2\theta}. 
\label{angmatt}
\ee

In matter both the eigenstates and the eigenvalues, and consequently, 
the mixing angle depend on matter density and neutrino energy. 
It is this dependence activates new degrees of freedom of the system and 
leads to qualitatively new effects. \\


{\it Resonance. Level crossing.}
According to (\ref{angmatt}), the dependence of the effective mixing parameter in 
matter,  $\sin^2 2\theta_m$, on density, neutrino energy as well as  the ratio of 
the 
oscillation and refraction lengths: 
\be
x \equiv \frac{l_{\nu}}{l_0} = \frac{2 E V}{\Delta m^2} \propto E n_e
\ee
has a resonance character. At 
\be
l_{\nu} = l_0 \cos 2\theta~~~~~~~{\rm  (resonance~~ condition)} 
\label{res}
\ee
the mixing becomes maximal: $\sin^2 2\theta_m = 1$. 
For small vacuum mixing the condition (\ref{res}) reads: 
\be
Oscillation~~ length \hskip 0.2cm \approx \hskip 0.2cm Refraction~~ length. 
\label{reseq}
\ee
That is, the eigen-frequency which characterizes a  system of mixed neutrinos, 
$1/l_{\nu}$,  coincides with the eigen-frequency of medium, $1/l_0$. 
 
For large vacuum mixing  ($\cos 2\theta_{12} = 0.4 - 0.5$) there is a 
significant  deviation from the equality (\ref{reseq}). Large vacuum mixing 
corresponds 
to the case of strongly coupled system  for 
which the shift of frequencies occurs. 

The resonance condition  (\ref{res}) determines the resonance density: 
\be
n_e^{R} = \frac{\Delta m^2}{2E} \frac{\cos 2\theta}{\sqrt{2} G_F}~.
\label{eq:resonance}
\ee
The width of resonance on the half of the height (in the density scale) is given by 
\be
2 \Delta n_e^R = 2 n_e^R \tan 2\theta .  
\label{width}
\ee
Similarly, one can introduce the resonance energy and the width of  
resonance in the energy scale. 

In medium with varying density, the layer where 
the density changes in the interval 
\be
n_e^R \pm \Delta n_e^R   
\ee
is called the resonance layer. 
In  resonance,  the level splitting  (difference of the 
eigenstates $H_{2m} - H_{1m}$ ) is minimal \cite{cab,bet}  and 
therefore the oscillation length being inversely 
proportional the level spitting,  is maximal. 

\subsection{Soft neutrino masses}

One possible deviation  from the standard scenario
can be related to  existence of the ``soft neutrino'' masses
or situation when a part of neutrino masses are soft.
The  neutrino masses can be generated by some low energy physics, 
so that the masses change with energy (distance) scale;  
also environment effect on the masses becomes substantial.
Recently, such a possibility has been considered in the context of
MaVaN scenario \cite{mavan}. 
Neutrino mass is some function of a 
small VEV, $v$  of some new scalar field
$m_{\nu} = m_{\nu} (v)$, and in turn, $v$ 
can depend on an environment, and in particular, on
density of the background neutrinos ({\it e.g.} relic neutrinos).
Another possibility 
is that the effective neutrino mass
is generated by the exchange of light scalar boson that
couples with usual matter (leptons and quarks).
Scalar interactions lead to chirality-flip and
therefore to generation ot true mass (and not just change of the
dispersion relation as in the case of refraction.)
Denoting the coupling constants of scalar boson with
neutrinos and  charged fermions  by
$\lambda_\nu$ and $\lambda_f$ correspondingly, 
we find  the soft mass
\be
m_{soft} = \frac{\lambda_\nu \lambda_f n_f}{m_\phi^2}. 
\ee
So,  in the evolution equation that describes
oscillations one has
\be
m_{\nu} = m_{vac} + m_{soft}, 
\ee
where $m_{vac}$ is a mass generated by some short range physics, 
{\it e.g.},  the electroweak scale VEV. 

\subsection{Degrees of freedom}

An arbitrary neutrino state can be expressed in terms of the
instantaneous eigenstates of the Hamiltonian, $\nu_{1m}$ and  $\nu_{2m}$, 
as
\be
\nu (t) = \cos\theta_a \nu_{1m} + \sin \theta_a \nu_{2m} e^{i\phi}~,
\label{exp}
\ee
where
\begin{itemize}

\item
$\theta_a = \theta_a (t)$ determines the  admixtures of  
eigenstates in $\nu (t)$;\\

\item
$\phi(t)$ is the phase difference between the two eigenstates (phase
of oscillations):
\be
\phi(t) = \int_0^t \Delta H dt' + \phi(t)_T~,
\ee  
here $\Delta H \equiv H_{1m} - H_{2m}$. 
The integral gives the adiabatic phase
and $\phi(t)_T$  can be related to
violation of adiabaticity. It may also have a
topological contribution  (Berry phase) in more complicated systems;\\

\item 
$\theta_m(n_e(t))$ determines the  flavor content of the eigenstates:   
$\langle \nu_e| \nu_{1m}\rangle = \cos \theta_m $, {\it etc.}.

\end{itemize}

Different processes are associated with these three 
different degrees of freedom.

\subsection{Oscillations in matter. Resonance enhancement of oscillations}

In medium with constant density the mixing is constant: 
$\theta_m (E, n) = const$. Therefore

\begin{itemize}

\item
the flavors of the eigenstates do not change; 

\item
the admixtures of the eigenstates do not change;  
there is no $\nu_{1m} \leftrightarrow \nu_{2m}$ transitions, 
$\nu_{1m}$ and  $\nu_{2m}$ are the eigenstates of propagation; 

\item
monotonous increase of the phase difference between the eigenstates occurs: 
$\Delta \phi_m = (H_{2m} - H_{1m}) t$. 

\end{itemize}

This is  similar to what happens in vacuum.  
The only operative degree of freedom is the phase. Therefore,  as in  vacuum, 
the evolution of neutrino  has a character of oscillations. 
However,  values of the oscillation parameters  (length, depth)  differ from those 
in vacuum. 
They are determined by the mixing in matter and by the effective energy splitting in matter: 
\be
\sin^2 2\theta   \rightarrow     \sin^2 2\theta_m, ~~~~~~  
l_{\nu} \rightarrow  l_{m} = \frac{2\pi}{H_{2m} - H_{1m}}. 
\ee

For a given density of matter the parameters of oscillations depend on the neutrino energy 
which leads to characteristic  modification of the energy spectra. 
Suppose a source produces the $\nu_e$- flux $F_0(E)$. 
The flux crosses a layer of length, $L$,  
with a constant density $n_e$  
and then detector measures the electron component of the flux 
at the exit from the layer, $F (E)$. 
\begin{figure}[htb]
\centering
\includegraphics[width=70mm,angle=000]{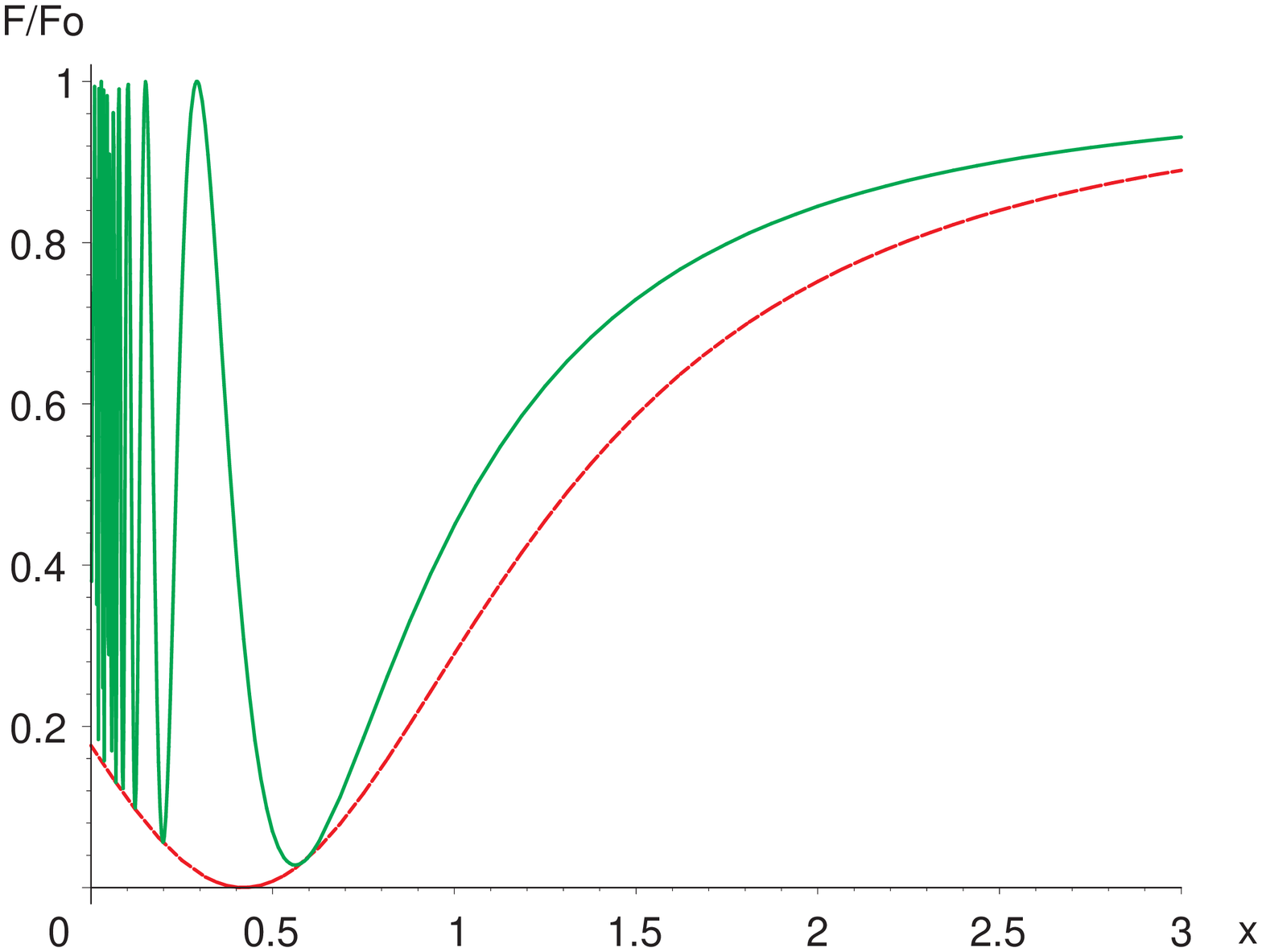}\\ 
\includegraphics[width=70mm,angle=000]{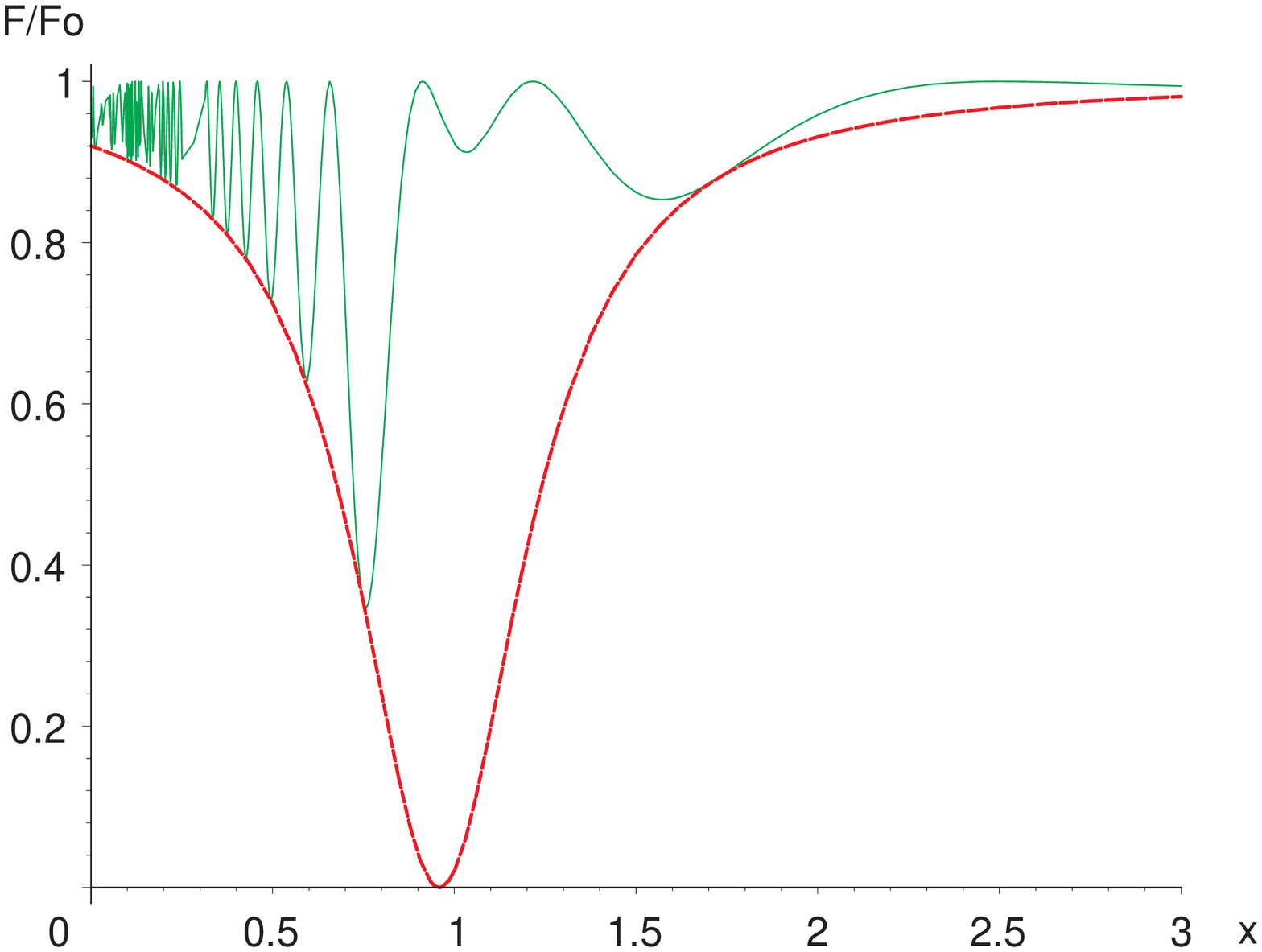}
\caption{~~Resonance enhancement of oscillations in matter with constant density.
Shown is a dependence of the ratio of the final and original fluxes, 
$F/F_0$,  on energy ($x \equiv l_\nu/l_0  
\propto E$) for a thin layer, $L = l_0/\pi$ (left panel)   and 
thick 
layer $L =  10 l_0/ \pi$ (right panel). $l_0$ is the refraction length. 
The vacuum mixing equals $\sin^2 2\theta = 0.824$.  }
\label{resenh}
\end{figure}
In fig.~\ref{resenh} we show  dependence of  the ratio $F(E)/F_0(E)$ 
on energy for thick and thin layers. 
The oscillatory curve is inscribed in to the resonance curve $(1 - \sin^2 2\theta_m)$. 
The frequency  of the oscillations  
increases with the length $L$. At the resonance energy, 
the oscillations proceed with maximal depths.  
Oscillations are enhanced in the resonance range:  
\be
E = E_R \pm \Delta E_R, ~~~ \Delta E_R = E_R \tan 2\theta  = E_R^0 \sin 2\theta, 
\ee 
where $E_R^0 = \Delta m^2/2\sqrt{2} G_F n_e$.  
Notice that for $E \gg E_R$,  matter suppresses the   oscillation depth;  
for small mixing the resonance layer is 
narrow, and the oscillation length in the resonance is large.  
With increase of the vacuum mixing: $E_R \rightarrow 0$ and $\Delta E_R \rightarrow  E_R^0$. 

The oscillations in medium with nearly constant density 
are realized for neutrinos of different origins crossing the mantle of the Earth. 

\subsection{MSW: adiabatic conversion}

In non-uniform medium, density changes on the way of neutrinos: 
$n_e = n_e(t)$. Correspondingly, the Hamiltonian of  system depends on time,  
$H = H(t)$, and  therefore, 

(i) the mixing angle changes during  propagation: 
$\theta_m = \theta_m (n_e(t))$;

(ii) the (instantaneous) eigenstates of the Hamiltonian, 
$\nu_{1m}$ and  $\nu_{2m}$,  are no more the 
``eigenstates" of propagation: 
the transitions $\nu_{1m} \leftrightarrow \nu_{2m}$ occur. 

However, if the density changes slowly enough 
the transitions 
$\nu_{1m} \leftrightarrow \nu_{2m}$ can be neglected. 
This is the essence of the 
adiabatic condition: $\nu_{1m}$ and $\nu_{2m}$ propagate 
independently,  as in  vacuum or  uniform medium. \\

{\it Evolution equation for the eigenstates. Adiabaticity. }
Let us consider the  adiabaticity condition. 
If external conditions (density) change slowly,  
the system (mixed neutrinos) has time to adjust this change. 

To formulate this condition 
let us consider the evolution equation for the 
eigenstate of the Hamiltonian in matter. 
Inserting  $\nu_f = U (\theta_m) \nu_m$ in to equation for the  
flavor states (\ref{eqmatt2}) we obtain 
\be
i\frac{d\nu_m}{dt} =  
\left(\begin{array}{cc}
H_{1m}  &  - i \dot{\theta}_m \\
i \dot{\theta}_m  &  H_{2m}  
\end{array}
\right)\nu_m .  
\label{eqmatt2a}
\ee
As follows from this equation for the neutrino eigenstates \cite{ms1,mess}, 
$|\dot{\theta}_m|$ determines the energy of transition 
$\nu_{1m} \leftrightarrow \nu_{2m}$ and  $|H_{2m}  - H_{1m}|$ 
gives the energy gap between levels. 

If \cite{mess}
\be
\gamma = \left| \frac{\dot{\theta}_m}{H_{2m}  - H_{1m}} \right| \ll 1, 
\label{adiab}
\ee
the off-diagonal terms can be neglected and system of equations for the 
eigenstates split. 
The condition (\ref{adiab}) means that 
the transitions $\nu_{1m} \leftrightarrow \nu_{2m}$ can be neglected 
and the eigenstates propagate independently 
(the angle  $\theta_a$  (\ref{exp}) is constant). 

For small mixing angles the adiabaticity condition is crucial in the resonance layer 
where  (i) the level splitting is small and (ii) the mixing angle changes rapidly. 
If the vacuum mixing is small, the adiabaticity is the most critical in the resonance point. 
It takes the form \cite{ms1}
\be
\Delta r_R > l_R, 
\label{adiab2}
\ee
where $l_R = l_{\nu}/\sin 2\theta$ is the oscillation length in resonance, 
and $\Delta r_R = $   
$n_R/ (dn_e/dr)_R \tan 2\theta$ is the spatial width of 
resonance 
layer. \\

{\it MSW- effect}. Dynamical features of the adiabatic 
evolution can be summarized in the following way: 

\begin{itemize}
 
\item
The flavors of the eigenstates change according to density change. 
The flavor composition of the eigenstates is determined by $\theta_m(t)$. 

\item
The admixtures of the eigenstates in a propagating neutrino state do not change 
(adiabaticity: no $\nu_{1m} \leftrightarrow \nu_{2m}$ transitions). The admixtures
are given by the  mixing in the  production point, $\theta_m^0$. 

\item
The phase difference increases; 
the  phase velocity is determined by the level splitting  
(which in turn,  changes with density (time)).  

\end{itemize}

Now two degrees of freedom become operative: the relative phase and 
the flavors of neutrino eigenstates. 
The MSW effect is  driven by the  change of  flavors of the  neutrino 
eigenstates in matter with varying density.  
The change of phase produces the  oscillation effect 
on the top of the adiabatic conversion.\\

Let us derive the adiabatic formula \cite{ms1,bet,par,hax}.  
Suppose in the initial moment the state $\nu_e$ is produced 
in matter with density $n_0$. Then the neutrino state can be written 
in terms of the eigenstates in matter as 
\be
|\nu_i \rangle  = |\nu_e \rangle  =  \cos \theta_m^0 
| \nu_{1m} \rangle + 
\sin \theta_m^0 |\nu_{2m} \rangle, 
\label{inst}
\ee
where $\theta_m^0 = \theta_m  (n_0)$ is the mixing angle in matter in the 
production point. 
Suppose this state propagates adiabatically to the region with 
zero density (as it happens in the case of the Sun).    
Then, the adiabatic evolution 
will consists of transitions $\nu_{1m} \rightarrow  \nu_{1}$, 
$\nu_{2m} \rightarrow  \nu_{2}$, and no transition between the eigenstates occurs,  
so the admixtures are conserved. As a result the final 
state  is
\be
|\nu (t) \rangle  =   \cos \theta_m^0 |\nu_{1} \rangle  + 
\sin \theta_m^0  e^{i\phi(t)} |\nu_{2} \rangle  , 
\label{finst}
\ee
where $\phi$ is the adiabatic phase. The survival probability 
is then given by 
\be
P = |\langle \nu_e| \nu (t) \rangle|^2. 
\ee    
Plugging $|\nu (t)\rangle $ (\ref{finst}) and $|\nu_e \rangle$  given by (\ref{2nu}) into this 
expression and performing  averaging 
over the phase which means that the contributions from 
$|\nu_1 \rangle $ and $|\nu_2 \rangle $ add incoherently,  we obtain 
\bea
P = (\cos \theta \cos \theta_m^0)^2 + (\sin \theta \sin \theta_m^0)^2 
\nonumber \\
= \sin^2 \theta  +  \cos 2 \theta \cos^2 \theta_m^0. 
\label{adiabform}
\eea
This formula gives description of the solar neutrino conversion with 
accuracy $10^{-7}$, that is,   corrections 
due to the adiabaticity violation are extremely small \cite{HLS}. \\

{\it Physical picture of the adiabatic conversion.}  
According to the dynamical  conditions, the admixtures of 
eigenstates are determined by the 
mixing in  neutrino production point. This mixing in turn, 
depends on the density in the initial point,  
$n_e^0$,  as compared to the resonance density. 
Consequently, a picture of the conversion depends on 
how far from the resonance layer (in the density scale) a neutrino is produced. 

\begin{figure}[t]
\centering
\includegraphics[width=80mm,angle=000]{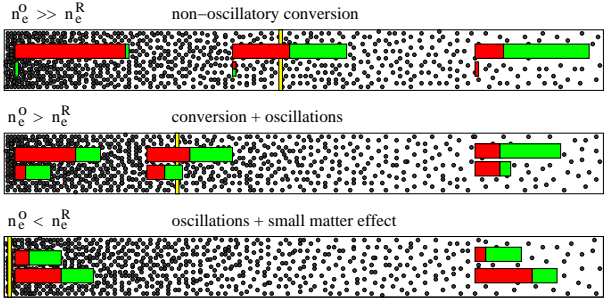}
\caption{~~Adiabatic  evolution of  neutrino state for 
three different initial condition ($n_e^0$). 
Shown are the neutrino states  in different moments of propagation in medium 
with varying (decreasing) density. The yellow vertical line indicates 
position of resonance. The initial state is $\nu_e$ in all the cases.  
The sizes of the boxes do not change, 
whereas the flavors (colors) follow the density change. 
}
\label{faev}
\end{figure}

Three possibilities relevant for solar neutrino conversion     
are shown in fig.~\ref{faev}.     
The state produced as $\nu_e$ propagates 
from large density region  to zero density. 
Due to adiabaticity  the sizes of  boxes which correspond 
to the neutrino eigenstates do not change. 

1).  $n_e^0 \gg n_e^R$: production far above the resonance (the upper panel).  
The initial mixing is strongly suppressed, 
and consequently,   the neutrino state, $\nu_e$,  consists mainly of one 
($\nu_{2m}$)  eigenstate, and furthermore, one flavor dominates in this   
eigenstate.  In the resonance 
(its position is marked by the yellow line) the mixing is maximal: 
both flavors are present equally. 
Since the admixture of the second eigenstate is very small, oscillations 
(interference effects) are strongly suppressed. 
So, here we deal with  the non-oscillatory flavor transition 
when the flavor of whole state (which nearly coincides with 
$\nu_{2m}$) follows the density change. 
At zero density we have $\nu_{2m} = \nu_{2}$,  and therefore 
the probability to find the electron neutrino (survival probability) equals
\be
P = |\langle \nu_e | \nu(t) \rangle|^2 \approx |\langle \nu_e |\nu_{2m}(t) \rangle|^2
= |\langle \nu_e |\nu_{2} \rangle|^2 \approx \sin^2 \theta. 
\label{surv1}
\ee
This result corresponds to $\theta^0_m = \pi /2$ in formula (\ref{adiabform}). 

The value of final probability, $\sin^2 \theta$, is the feature 
of the non-oscillatory transition. 
Deviation from this value indicates a presence of oscillations. 

2).  $n_e^0 > n_e^R$:  production above the resonance (middle panel).
The initial mixing is not suppressed.  Although $\nu_{2m}$ is the main component, 
the second eigenstate, $\nu_{1m}$,  
has an appreciable  admixture; also    
the flavor mixing  in the neutrino eigenstates is significant. 
So, the  interference effect is not suppressed. 
As a result,  here an interplay of the adiabatic conversion and 
oscillations occurs. 

3). $n_e^0 < n_e^R$: production below the resonance (lower panel). 
There is no  crossing of the resonance region.  
In this case the matter effect  gives only corrections to the vacuum oscillation picture.

The resonance density is inversely proportional to the 
neutrino energy: 
$n_e^R \propto 1/E$. So, for the same density profile, the condition 
1) is realized for high energies, the condition 2) is realized for  
intermediate  energies and  condition 3) -- for low energies. As we will see all three case 
realize for solar neutrinos. \\

The adiabatic transformations show universality:  The averaged probability 
and the depth of oscillations in a given moment 
of propagation are determined by the density in a given point and by 
initial condition (initial density and flavor). They do not depend on 
density distribution between the initial and final points.  
In contrast, the phase of oscillations is an 
integral effect of  previous evolution and it depends on 
a density distribution. 

The universal character of the adiabatic conversion can be further generalized in terms of 
variable~\cite{ms1}  
\be
y \equiv \frac{n_e^R - n_e}{\Delta n_e^R}
\label{y-var}
\ee
which is the distance (in the density scale) from the resonance density  in the units of 
the width of resonance layer (\ref{width}). In terms of $n$ the conversion pattern 
depends only on the initial value $y_0$.

\begin{figure}[thb]
\includegraphics[width=70mm,angle=000]{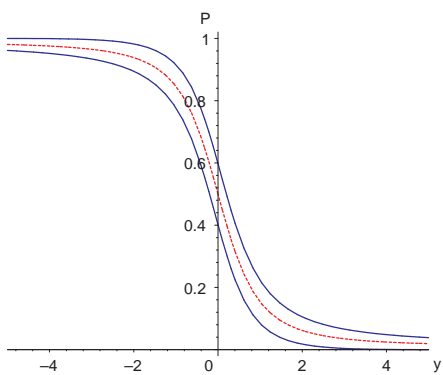}
\caption{The dependence of the average probability (dashed line) and the depth of oscillations
($P^{max}$, $P^{min}$ solid lines) on $y$  for  $y_0 = - 5$.
The resonance layer corresponds to $y = 0$. For $\tan^2 \theta = 0.4$ (large mixing MSW solution)
the evolution stops at $y_f = 0.47$.
}
\label{msw}
\end{figure}

In fig.~\ref{msw}   we show  dependences 
of the average probability, $\bar{P}$, and depth of oscillations  determined by    
$P^{max}$ and $P^{min}$,  on $y$. The probability itself is the oscillatory function 
which  is inscribed into the band shown by the solid lines. 
The average probability is shown by the  dashed 
line.   The curves are determined by the initial value 
$y_0$ only. In particular,  there is no explicit dependence on the vacuum mixing angle. 
The resonance is at $y = 0$ and  the resonance layer is given by the interval 
$y = -1 \div  1$. The figure corresponds to $y_0 = - 5$,  {\it   i.e.}, to production  
above the resonance layer; the oscillation depth is relatively small.  
With further decrease of  $y_0$, the oscillation band  becomes narrower 
approaching the line of  non-oscillatory conversion. 
For zero final density we have 
\be
y_f = \frac{1}{\tan 2\theta}. 
\ee
So,  the vacuum mixing enters the final condition. 
For the best fit LMA point,  $y_f = 0.45 - 0.50$, and  the evolution should stop 
rather close top the resonance. 
The smaller mixing the larger final $y_f$ and the stronger transition. 

\subsection{Adiabaticity violation}

In the adiabatic regime the probability 
of transition between the eigenstates is exponentially 
suppressed  $P_{12} \sim exp{(- \pi/2\gamma)}$ 
and $\gamma$ is given in (\ref{adiab}) \cite{hax,par}. 
One can consider such a transition as penetration  through a barrier of the height 
$H_{2m} - H_{1m}$  by a  system  with the kinetic energy $d\theta_m /dt$.

If density changes rapidly, so that the condition (\ref{adiab}) is not satisfied, 
the transitions $\nu_{1m} \leftrightarrow \nu_{2m}$ become efficient. 
Therefore  the admixtures of the eigenstates in a given propagating state 
change. In our pictorial representation (fig. 3) the sizes of boxes change. 
Now all three degrees of freedom of the system become operative.  

Typically, adiabaticity breaking leads to weakening of the  flavor  
transition.  The non-adiabatic transitions can
be  realized inside  supernovas for the very small 1-3 mixing.

\section{Determination of the oscillation parameters}

\subsection{Solar neutrinos}

{\it Data.} Data analysis is based on  results from the
Homestake experiment~\cite{homestake}, 
Kamiokande and SuperKamiokande~\cite{sol05}, 
from radiochemical Gallium experiments 
SAGE~\cite{sage}, Gallex~\cite{gallex} and GNO~\cite{gno} and from SNO~\cite{sno}. 
The information we have collected can be described in three-dimensional 
space: 

1.  Type of events: $\nu e$ scattering (SK, SNO), 
CC-events (Cl, Ga, SNO) and NC events (SNO).  

2.  Energy of events: radiochemical experiments integrate 
effect over the energy from the threshold to  
the maximal energy in the spectrum. Also NC events are integrated over energies. 
The CC events in  SNO and  $\nu e$ events at 
SuperKamiokande give information about the energy  
spectrum  of original neutrinos.  

3. Time dependence of rates (searches for time variation of the flux). \\

{\it Evidence of conversion.}
There are three types of observations which testify for the 
neutrino conversion: 

1). Deficit of signal which implies the deficit of the 
electron neutrino flux. 
It can be described by the ratio 
$R \equiv N^{obs}/N^{SSM}$, where $N^{SSM}$ is the 
signal predicted according to the Standard solar model fluxes  
\cite{john}. 
The deficit has been found in all (but SNO neutral current)
experiments. 

2). Energy spectrum distortion - dependence of the 
suppression factor on energy. Indirect evidence is provided 
by comparison of the deficits  in experiments sensitive to 
different energy intervals:  
\bea
Low ~energies ~ (Ga):~~~  R = 0.5 - 0.6\\
High ~energies~ (Cl,~ SK,~ SNO): ~~~ R \approx 0.3. 
\eea
So, the deficit  increases with  neutrino energy.

3). Smallness of ratio of signals due to charged currents 
and neutral currents~\cite{sno}: 
\be
\frac{CC}{NC} = 0.340 \pm 0.023~(stat.) ~^{+ 0.029}_{-0.031}(syst.) ~.
\ee 
The latter is considered as the direct evidence
of the flavor conversion since NC events are not 
affected by this conversion, whereas the number CC events 
is suppressed.

All this testifies for the LMA MSW solution.

Till now there is no statistically significant observations of other 
signatures of the conversion, namely,   

- distortion of the boron neutrino spectrum: up turn at low energies  
in SK and SNO  
(significant effect should be seen  below 5 - 7 MeV); 

- day-night effect (recall that SK agrees with predictions however 
significance is about $1\sigma$);

- semiannual time variations  on the top of annual 
variations (due to eccentricity of the Earth orbit).\\

{\it Physics of conversion} \cite{HSsol}.
Physics can be described in terms of three effects: 

1). Adiabatic conversion (inside the Sun);

2). Loss of coherence of the neutrino state (on the way to the Earth);

3). Oscillations of the neutrino mass states in the matter of the Earth.

According to LMA,  inside the Sun the initially 
produced electron neutrinos 
undergo the highly  adiabatic conversion: $\nu_e \rightarrow \cos\theta_m^0 \nu_1 
+ \sin\theta_m^0 
\nu_2$, where $\theta_m^0$ is the mixing angle in the production point.  
On the way from the central parts of the Sun the coherence of  
neutrino state is lost after several hundreds oscillation 
lengths~\cite{HSsol}, and  incoherent fluxes 
of the mass states $\nu_1$ and $\nu_2$  arrive
at the surface of the Earth. In the matter of the Earth $\nu_1$ and $\nu_2$ 
oscillate partially regenerating the $\nu_e$-flux. 
With regeneration effects included 
the averaged survival probability can be written as 
\be
P = \sin^2 \theta +  \cos^2 \theta_{12}^{m0} \cos 2\theta_{12} - 
\cos 2\theta_{12}^{m0} f_{reg}.  
\label{surv}
\ee
Here the first term corresponds to the non-oscillatory transition (dominates  
at the high energies), the second term is the contribution from the 
averaged oscillations which increases with decrease of  energy,  and the 
third term is the regeneration effect, with the regeneration factor, 
$f_{reg}$ defined as   
\be 
f_{reg} \equiv P_{2e} - \sin^2 \theta.  
\ee 
Here $P_{2e}$ is the probability of  
$\nu_2 \rightarrow \nu_e$ transition in the matter of the Earth 
(without oscillations in matter: $P_{2e} = \sin^2 \theta$). 
At low energies $P$ reduces to the vacuum 
oscillation probability with very small matter corrections.  

\noindent
There are  three energy ranges with different features of transition: 

1. In the high energy part of spectrum, $E > 10$ MeV ($x \equiv l_{\nu}/l_0 > 2$), 
the  adiabatic conversion with small  oscillation effect occurs. 
At the exit, the resulting averaged probability 
is slightly larger than $\sin^2 \theta$ expected  from  the non-oscillatory transition.  
With decrease of energy the initial density approaches the resonance density, 
and the depths of oscillations increases. 

2. Intermediate energy range $E \sim (2 - 10)$ MeV ($x = 0.3 - 2$ ) 
the oscillation effect is significant. 
The interplay of the oscillations and conversion takes place.

For $E \sim 2$ MeV  neutrinos are produced in resonance.  

3. At low energies: $E <  2$ MeV ($x < 0.3$), 
the  vacuum oscillations  with small matter corrections occur. 
The averaged survival probability $P \approx 0.5 \sin^2 2 \theta$ is given by 
approximately the vacuum oscillation formula. \\ 



{\it Inside the Earth.} 
Entering the Earth the state $\nu_{2}$ (which dominates at high energies) splits in two 
matter eigenstates: 
\be
\nu_{2} \rightarrow \cos\theta_m'  \nu_{2m} + \sin \theta_m'\nu_{1m}.
\ee  
It oscillates regenerating partly the  $\nu_e$-flux. 
In the approximation of 
constant density profile the regeneration factor  equals 
\be
f_{reg} = 
0.5 \frac{l_{\nu}}{l_0}~\sin^2 2\theta =  
\frac{EV}{\Delta m^2}~\sin^2 2\theta.  
\label{freg}
\ee
Notice that the oscillations of $\nu_{2}$  
are  pure matter effect and for the  presently favored 
value of $\Delta m^2$ this effect is small. According to (\ref{freg}), 
$f_{reg} \propto 1/\Delta m^2$ and 
the expected day-night asymmetry  of the charged current signal 
equals
\be
A_{DN} = f_{reg}/P  \sim (3 - 5)\%~. 
\label{freg1}
\ee

Apparently the Earth density profile is not constant and it consists 
of several layers with slow density change  and jumps 
of density on the borders between layers.  
It happens that for solar neutrinos 
one can get simple analytical result  for oscillation probability for 
realistic density profile. 
Indeed, the solar neutrino oscillations occur in the so called low energy regime 
when
\be
\epsilon \equiv  \frac{2EV(x)}{\Delta m^2} \ll 1, 
\label{earth}
\ee
which means that the potential energy 
is much smaller than the kinetic energy.  
For the LMA oscillation parameters and the solar  
neutrinos:   $\epsilon(x) = (1 - 3)\cdot 10^{-2}$.  
In this case one can use small parameter 
$\epsilon(x)$ (\ref{earth}) to develop the perturbation theory 
\cite{ara1}. 
The following expression for the  regeneration factor,   
$f_{reg}$, has been obtained \cite{ara1,atv}
\be
f_{reg} = 
\frac{1}{2} \sin^2 2\theta \int_{x_0}^{x_f}dx V(x) \sin \phi_m (x \rightarrow 
x_f).
\label{reg}
\ee
Here $x_0$ and $x_f$ are the initial 
and final points of propagation correspondingly,  
and 
$\phi_m (x \rightarrow x_f)$
is the adiabatic phase  acquired 
between a given point of trajectory, $x$, and final point, $x_f$. 
The latter feature has important consequence leading to the attenuation effect - 
weak sensitivity to the remote structures of the density profile when non-zero energy 
resolution of detector is taken into account.  On the other hand $f_{reg}$
can be strongly affected by some relatively small structures near the surface of the Earth.

Another insight into phenomena can be obtained using the adiabatic perturbation 
theory which leads to~\cite{HLS}   
\be 
f_{reg} = \frac{2E \sin^2 2\theta}{\Delta m^2} \sin \frac{\phi_0}{2} 
\sum_{j = 0... n-1} \Delta V_j \sin \frac{\phi_j}{2}. 
\label{sfum}
\ee    
Here $\phi_0$ and $\phi_j$ are the phases acquired 
along whole trajectory and on the part of the trajectory 
inside the borders $j$. This formula corresponds to  
symmetric profile with respect to the center of trajectory. 
Using (\ref{sfum}) one can easily infer the attenuation effect. 
The formula reproduces precisely the results of exact 
numerical calculations. 
Notice that the adiabatic perturbation theory is relevant here 
because the adiabaticity is fulfilled within the layers and 
maximally broken at the borders.\\ 

{\it Determination of the solar oscillation parameters.} 
Knowledge of the energy dependence of the adiabatic conversion allows 
one to connect the oscillation parameters with observables immediately.

1). Determination of the mixing angle. 
To explain stronger deficit at higher energies one needs to 
have $\theta < \pi/4$ or $\sin^2 \theta < 1/2$. 
Furthermore, using the fact that  
$P_h > \sin^2 \theta$ and $P_l < 0.5 \sin^2 2\theta$ 
we find  
\be
\frac{P_h}{P_l} \geq \frac{\sin^2 \theta}{0.5 \sin^2 2\theta}
=\frac{1}{2\cos^2\theta}, 
\ee
where on the RHS we have taken the asymptotic values of the 
survival probability. Consequently, 
\be 
\sin^2 \theta \leq 1 - \frac{P_l}{2 P_h}  \sim 0.1 \pm 0.2. 
\ee

The ratio of CC to NC events determines the survival probability:  
\be
P = \sin^2 \theta  + \cos2 \theta \langle \cos^2 \theta_m^0 \rangle = \frac{CC}{NC}.
\ee
For high energies and without Earth matter regeneration effect 
$P = \sin^2 \theta$. Since no significant distortion 
of the energy spectrum is seen at SK and SNO  
the Boron neutrino spectrum  should be in the flat part (bottom of the 
``suppression pit''). 
In this region the deviation from asymptotic value is weak. 
For $\Delta m^2 \sim 8 \cdot 10^{-5}$ eV$^2$ the 
averaged oscillation effect is about $10\%$. Therefore 
\be
\sin^2 \theta_{12} \approx 0.9 \frac{CC}{NC} \approx 0.31.
\ee
  
2). Determination of $\Delta m^2$. 
Value of suppression in the Gallium experiments,  $P_l$,  implies that the 
$pp$- spectrum is in the vacuum dominated region, whereas  
stronger suppression of SK and SNO signals (together with 
an absence of distortion) means that the boron neutrino flux 
is in the matter dominated region. So,  the transition region 
should be  $E_{tr} \sim (1 - 4)$ MeV. 
On the other hand the expression for the middle energy 
of the transition region equals (it corresponds to neutrino production in resonance)
\be
E_{tr} = \frac{\Delta m^2 \cos 2\theta}{2V_{prod}}, 
\label{e-tr}
\ee
where $V_{prod}$ is typical potential in the neutrino 
production region in the Sun.  From (\ref{e-tr}) we obtain 
\be
\Delta m^2 = \frac{2 E_{tr}V_{prod}}{\cos 2\theta}
\label{m-12}
\ee
which gives $\Delta m^2 = (3 - 15) \cdot 10^{-5} ~~{\rm eV}^2$ 
in the correct range.  

Another way to measure $\Delta m^2$ is to study  the high energy effects:
according to LMA the splitting $\Delta m^2$ is restricted from below by 
the increasing 
day-night asymmetry and from above by absence of the significant 
up turn of spectrum at low energies. \\


New SNO results are expected from
the third (last) phase of the experiment that employs
the $^3$He-counters for neutrons. The counters 
provide with a better identification of the NC-events 
and therefore preciser measurements  of the  CC/NC ratio, 
and  the  $\theta_{12}$ angle (combination of
$\cos^4\theta_{13}\sin^2 \theta_{12}$
in the three neutrino context). 
BOREXINO should start measurements soon \cite{borexino}. 

The SAGE calibration result is about $2\sigma$ below
expectation \cite{calibr}. That may testify for lower cross-section and therefore
higher $pp-$flux at the earth due to larger survival probability.
That produces some tension in the fit of the solar data \cite{fogli06}.
Another possibility proposed recently is that the  reduced
calibration result  is due to short range oscillations
to sterile neutrinos \cite{giunti06}.

Searches for time variations and possible periodicity 
in the solar neutrino data are continued \cite{period}.

\subsection{KamLAND}

KamLAND (Kamioka Large Anti-neutrino detector) is the reactor long baseline 
experiment \cite{kamland} . Few relevant details: 1kton liquid scintillator detector 
situated in the Kamioka laboratory detects the antineutrinos from 
surrounding atomic reactors (about 53) with the effective distance 
(150 - 210) km. The classical reaction of the inverse beta decay,  
$\bar{\nu}_e p \rightarrow e^+ n$, is used. 
The data include 

(i) the total rate of events;

\begin{figure}[htb]
\centering
\includegraphics[width=70mm,angle=000]{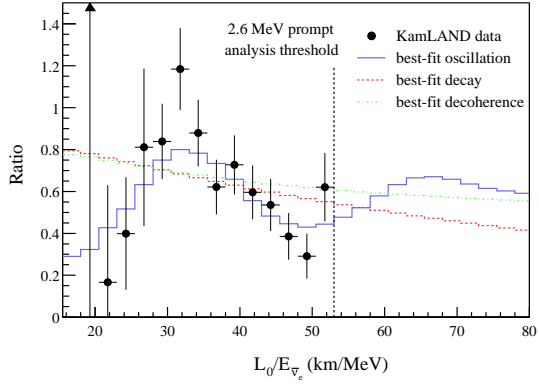}
\caption{The $L/E$ distribution of events in the KamLAND experiment; from 
\cite{kamland}.
}
\label{kamsp}
\end{figure}

(ii) the energy spectrum (fig. \ref{kamsp});

(iii) the time dependence of the signal which is due to 
variations of the reactors power. (Establishing the 
correlation between the neutrino signal and power of reactors 
is important check of the whole experiment). 
In fact, this change also influences the oscillation effect   
since the effective distance from the reactors 
changes ({\it e.g.}, when power of the closest  reactor decreases).

In the oscillation analysis the energy threshold $E > 2.6$ MeV 
is established. 

The physics process is essentially the vacuum oscillations of $\bar{\nu}_e$. 
The matter effect, about $1\%$,  is negligible at the present 
level of accuracy.  

The evidences of the oscillations are 

1). The deficit of the number of the $\bar{\nu}_e$ events 
\be 
R_{\nu} = \frac{N_{obs}}{N_{expect}} = \frac{258}{365.2 \pm 23.7}  \sim 0.7
\ee
for $E > 2.6$ MeV.

2). The distortion of the energy spectrum or $L/E$ dependence 
(when  some reactors switch off the effective 
distance changes). 
Notice that an absence of strong spectrum 
distortion excludes large part of the oscillation 
parameter space. 

Oscillation parameters are related to the observables 
in the following way. 
The main features of the $L/E$ dependence are   maximum at 
$(L/E)_{max} = 32$ km/MeV (phase $\phi = 2\pi $)
and minima at $L/E_m = 16,~ 48$ km/MeV ($\phi = \pi, 3\pi$). They  
fit well the  expected oscillation pattern. 
Taking the first maximum   we find 
$$
\Delta m^2 = \frac{4\pi}{(L/E)_{max}} = 8 \cdot 10^{-5}~~{\rm eV}^2.
$$

The deficit of the signal determines (for a given $\Delta m^2$) the 
value of mixing angle: 
\be 
\sin^2 2\theta_{12} = \frac{1 - R_{\nu}}{\langle \sin^2 \phi \rangle},  
\ee
where the averaged over the energy interval oscillatory factor
can be evaluated for  the KamLAND detector as 
$\langle\sin^2 \phi \rangle \sim 0.6$. 
Notice that sensitivity to mixing angle is not high at present. \\

Extracted values of the  oscillation parameters are  in a very good agreement with  
those obtained from the solar neutrino analysis. 
This comparison implies the CPT conservation. 

Combined analysis of the solar neutrino data and the 
KamLAND  can be performed in assumption of the CPT conservation.  
The mixing angle is mainly 
determined by the solar neutrino data, 
whereas  $\Delta m^2$  is fixed by the  KamLAND.
New complete calibration of the detector will allow
to improve sensitivity to 1-2 mixing \cite{KL-cal}.

Comparison of results from the solar neutrinos and KamLAND 
open important possibility to check the theory of neutrino 
oscillation and conversion,  test CPT, search for new neutrino interactions and 
new neutrino states.  








\subsection{Atmospheric neutrinos}

{\it Experimental results.} 
The atmospheric neutrino flux is produced in interactions of 
the high energy cosmic rays (protons, nuclei) with nuclei of  
atmosphere. The interactions occur at heights (10 - 20) km. 
At low energies the flux is formed in the chain of decays: 
$\pi \rightarrow \mu  \nu_{\mu}$, $\mu \rightarrow e   \nu_e  \nu_{\mu}$. 
So, each chain produces 
$2\nu_{\mu}$ and $1\nu_e$, and correspondingly, the ratio of fluxes equals 
\be
r \equiv \frac{F_{\mu}}{F_e} \approx 2.
\label{r-emu}
\ee
With increase of energy the ratio increases since the lifetime acquires 
the Lorentz boost and muons have no time to decay before collisions: 
they are absorbed or loose the energy. As a consequence, the flux of 
the electron neutrinos decreases. 


In spite of the long term efforts, still the predicted atmospheric neutrino fluxes 
have  large uncertainties (about 20\% in overall normalization and 
about 5\% in the so called ``tilt'' parameter 
which describes the uncertainty in the energy-dependence of the flux). 
The origin of uncertainties is twofold:  
original flux of the cosmic rays  
and  cross sections of interactions.  


The recent analyzes include the data from 
Baksan telescope, SuperKamiokande~ \cite{atm,tauapp}, 
MACRO \cite{macro}, 
SOUDAN~\cite{soudan}. 
The data can be presented in the three dimensional space 
which includes

- type of events detected: $e$-like events (showers), $\mu$-like events, 
multi-ring events, NC events (with detection of $\pi^0$), $\tau-$enriched  sample of events.

- energy of events: widely spread classification 
includes the  sub-GeV and  multi-GeV events,  
stopping muons, through-going muons, {\it etc.}.

- zenith angle (upward going, down going, {\it etc}).

Now MINOS experiment \cite{minos} provides  some  early information 
on oscillation effects for the atmospheric neutrinos and antineutrinos separately.\\

{\it The evidences of the atmospheric neutrino oscillations} include: 

1).  Smallness of the double ratio of numbers of $\mu$-like to $e$-like 
events~\cite{atm}: 
\be
R_{\mu/e} \equiv 
\frac{N_{\mu}^{obs}/N_{\mu}^{th}}{N_{e}^{obs}/N_{e}^{th}} . 
\label{doubler}
\ee 
The ratio weakly depends on energy: it slightly increases from  sub-GeV 
to  multi-GeV  range (as expected): 
\bea
R_{\mu/e} = 0.658 \pm 0.016 \pm 0.035 ~~{\rm (subGeV)}~~~~~~~~
\nonumber \\
R_{\mu/e} = 0.702 ^{+0.032}_{-0.030} \pm 0.101~~ {\rm (multiGeV+PC)}.
\eea  
Apparently in the absence of oscillations (or other non-standard neutrino 
processes) the double ratio should be 1.  
The smallness of the ratio testifies for disappearance of the 
$\nu_{\mu}-$flux. 

\begin{figure}[t]
\includegraphics[width=75mm,angle=000]{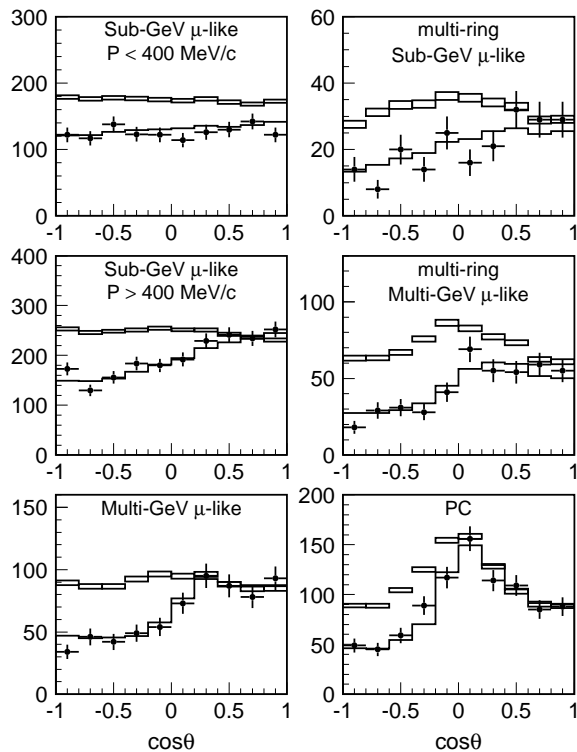}
\caption{The zenith angle distribution of the atmospheric
$\mu$-like events in different energy ranges; from \cite{atm}.}
\label{atzen}
\end{figure}

2). Distortion of the zenith angle dependence of the 
$\mu$ -like events (see fig. \ref{atzen}). 
Global characteristic of this distortion is 
the up-down asymmetry  defined as 
\be
A_{up/down} \equiv \frac{N_{up}}{N_{down}}. 
\ee 
Due to complete up-down symmetric configuration for the production,  
in the absence of oscillations or other non-standard effects the 
asymmetry should be absent: $A_{up/down} = 1$.  

The zenith angle dependence for different types of events in 
different ranges of energies is shown in fig. \ref{atzen} from \cite{atm}. 
The zenith angle of the neutrino trajectory 
is related to the baseline $L$ as 
$
L = D \cos\theta_z. 
$
So, studying the zenith angle distributions we study essentially 
the distance dependence of the oscillation probability. 

Substantial  distortion of the zenith angle 
distribution is found. The  deficit of numbers 
of events increases with decrease of $\cos \theta_Z$ 
and reaches about 1/2 in the upgoing vertical direction for multi-GeV 
events. The distortion increases with energy. 
Correspondingly,  the up-down asymmetry  increases with energy:  

In contrast to the $\mu$-like, the $e$-like events distribution 
does not show any anomaly. Though one can mark some excess  (about 15\%) 
of the $e$-like events in the sub-GeV range (upper-left panel of fig. \ref{atzen}). 

3). Appearance of the $\tau$-like events  ($2.4 \sigma$ effect) \cite{tauapp}. 

\begin{figure}[t]
\includegraphics[width=70mm,angle=000]{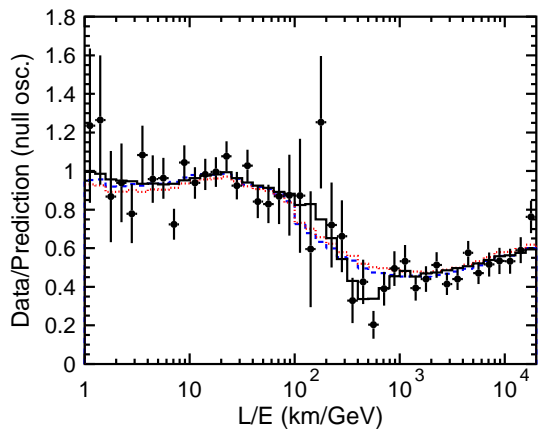}
\caption{$L/E$ distribution of the atmospheric $\mu$ like events; from \cite{leatm}. 
The solid line corresponds to the oscillation fit.}
\label{atdip}
\end{figure}

4). The $L/E-$dependence shows the first oscillation minimum (fig.~\ref{atdip}).\\
  
In the first approximation 
all these data can be consistently described in terms 
of the $\nu_{\mu} - \nu_\tau$ vacuum oscillations. 
Notice that for pure $2\nu$ oscillations of this type no matter effect 
is expected: the matter potentials of  $\nu_{\mu}$ and $\nu_\tau$  
are equal. In the context of three neutrino mixing, 
for non-zero values of $\sin \theta_{13}$ the matter effect
should be taken into account for the $\nu_{\mu} - \nu_\tau$ channel.  

As we marked above, the probabilities of $\nu_e$ and $\nu_{\mu}$ oscillations 
in matter of the Earth driven by 
the ``solar'' parameters  $\Delta m^2_{21}$ and $\sin^2 2\theta_{12}$ are large 
and even matter-enhanced in the sub-GeV range. However, observable effects  
of these oscillations are  suppressed by factor 
\be
(r \cos^2 \theta_{23} - 1),  
\ee 
where the ratio $r$ is defined in  eq. (\ref{r-emu}). 
In the sub-GeV range $r \approx 2$ and for maximal 2-3 mixing 
effects cancel. With increase of neutrino energy $r$ increases, 
however the probabilities are suppressed by matter effect. 

So, in the first approximation a unique description 
in terms of $\nu_{\mu} - \nu_\tau$ 
oscillation is valid for different types of 
events and in a very wide range of energies: from 0.1 to more than 100 GeV. \\  

{\it Determination of the atmospheric neutrino oscillation parameters.}
Let us describe how the oscillation parameters 
can be immediately related to  observables. 
We will use here the interpretation of the results 
in terms of $2\nu$-oscillations $\nu_\mu - \nu_\tau$. 

The most clean way to determine parameters is to use the 
zenith angle distribution of the multi-GeV $\mu$-like events. 
As follows from fig. \ref{atzen}
for the down-going muons, $\cos \theta_Z \sim 0.5 \div 1$,   the oscillation 
effects are negligible (good agreement with the no-oscillation 
predictions). For the up-going muons, 
$\cos \theta_Z \sim - 0.5 \div - 1$, 
there is already 
the averaging oscillation effect.  Transition region  
corresponds to  the 
horizontal events with  
$\cos \theta_Z \sim 0.0 \div 0.2$. For these events the 
baseline  $L = 500$ km  should be  comparable with  
the oscillation length $L \approx l_{\nu}$, so that  
\be 
\Delta m^2 = \frac{4\pi E_{multi-GeV}}{L_{horizon}}.
\ee
Taking $E = (1- 2)$ GeV 
we find $\Delta m^2 = (1-4) \cdot 10^{-3}$ eV$^2$.  
An  uncertainty in the neutrino direction 
and the fact that distance strongly depends on 
$\cos \theta_Z$ in the horizontal direction 
lead to the uncertainty in the determination of 
the atmospheric $\Delta m^2$.

For the upward-going $\mu$-like events the oscillations are averaged 
(no dependence of the suppression factor on $\cos \theta_Z$),  
so that $N^{obs}(up)/N^{th}(up) = 1 - \sin^2 2 \theta$. This allows us 
to determine the mixing angle: 
\be
\sin^2 2 \theta = 2[1 - N^{obs}(up)/N^{th}(up)].
\ee
$~$ From  the fig.~\ref{atzen}  $N^{obs}(up)/N^{th}(up) \approx 0.5$,  
and consequently,  $\sin^2 2 \theta = 1$ ($N^{th}(up) \approx N^{obs}(up)$).  

Other independent determinations are possible: 
in the sub-GeV range the zenith angle dependence is weak because 
of strong averaging effect: (i) the oscillation length is shorter 
and therefore the oscillations develop already 
for large part of the 
downgoing events; (ii) the angle between neutrino and detected 
muon is very large and directionality is essentially lost. 
So, taking the deficit  of the total number of events  we obtain 
\be
\sin^2 2 \theta \geq 2[1 - R],
\ee
where equality corresponds to  the developed oscillations 
for all directions. From fig.~\ref{atzen} we find $R = 0.67$, 
and therefore $\sin^2 2 \theta \geq 0.7$.    

To determine  mixing angle one can use also the double ratio. 
As follows from (\ref{doubler})
\be
\sin^2 2 \theta = \frac{1 - R_{\mu/e}}{\langle \sin^2 \phi/2\rangle_z}, 
\label{mixdoub}
\ee
where $\langle \sin^2 \phi/2\rangle_z$ is the averaged over 
the energy and zenith angle 
oscillatory factor.  For multi-GeV events 
$\langle \sin^2 \phi/2\rangle_z = 0.20 - 0.25$ and therefore 
from (\ref{mixdoub}) we obtain $\sin^2 2 \theta \sim 1$. 

The most precise determination of $\Delta m^2$ follows from the 
$L/E$ - dependence of the events (fig. \ref{atdip})  
which is  considered as the direct observation 
of the neutrino oscillations - oscillatory effect \cite{leatm}. 
In the first oscillation minimum - dip in the survival probability 
the phase of oscillations equals $\phi = \pi$. Therefore 
\be
\Delta m^2 = \frac{2\pi}{(L/E)_{dip}} . 
\label{lovere}
\ee
From fig. \ref{atdip}: $(L/E)_{dip} = 500$ km/GeV. This gives immediately 
$\Delta m^2 = 2.5 \cdot 10^{-3}$ eV$^2$.


Results of the  $3\nu-$analysis 
from \cite{concha} and \cite{bari} 
show one important systematic effect: 
the shift of 2-3 mixing from maximal one when the effect of the 
1-2 sector is included. Also whole allowed region is shifted. 
Even larger deviation of $\sin^2 \theta_{23}$ from 0.5  has been found in \cite{bari}. 
Essentially this result is related to the excess of the 
$e$-like events in the sub-GeV range.




\subsection{K2K}

The $\nu_{\mu}-$ beam with typical energies $E = (0.5 - 3)$ GeV 
was created at KEK and  directed to Kamioka.  
Its interations were detected by SuperKamiokande \cite{k2k}. 
The baseline  is about 250 km. 
The oscillations of muon neutrinos,  
$\nu_{\mu}  \rightarrow \nu_{\mu}$, 
have been studied by comparison of the detected number 
and the energy spectrum of the $\mu$-like events with the predicted ones. 
The predictions 
have been made by extrapolating the results from the ``front'' 
detector to the Kamioka place. The front detector similar to SK (but 
of smaller scale) was at  about 1 km distance from the 
source.  

\begin{figure}[htb]
\includegraphics[width=70mm,angle=000]{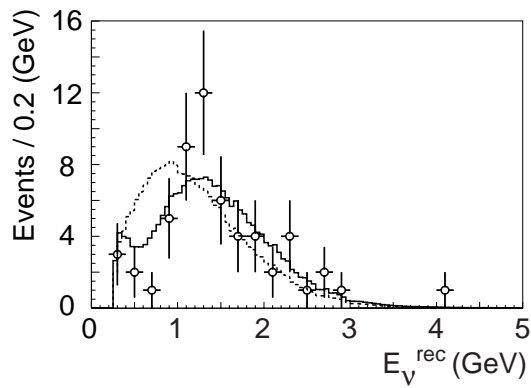}
\caption{The energy spectrum of events in the K2K experiment, from
\cite{k2k}. Also shown are equally normalized fit curves with oscillations (solid)
and without  oscillations (dotted)}
\label{spk2k}
\end{figure}

The evidences of oscillations were   
(i) the deficit of the total number of events: 
107 events have been observed whereas $151^{+12}_{-10}$ 
have been expected; (ii) the spectrum distortion 
(fig. \ref{spk2k}).  
Searches for  the $\nu_{\mu}  \rightarrow \nu_e$  
oscillations gave negative result. 

The data are interpreted as the non-averaged vacuum oscillations 
$\nu_{\mu} - \nu_{\tau}$. 

The energy distribution of the detected $\mu-$like events show 
an evidence of the first oscillation dip at $E \sim 0.5$ GeV 
(see fig. \ref{spk2k}). 
This allows one to evaluate  $\Delta m^2$.  
Using the relation (\ref{lovere}) with 
$L/E = 250{\rm km} / 0.5 {\rm GeV} = 500$ km/GeV 
(apparently the same as in the atmospheric neutrino case), we obtain 
$\Delta m^2 = 2.5 \cdot 10^{-3}$ eV$^2$ in perfect agreement with 
the atmospheric neutrino result. 
(In fact the data stronger exclude other values of $\Delta m^2$ 
than favor the best one.)

The substantial oscillation suppression is present in the 
low energy part of the spectrum  ($E < 1$  GeV) only. Therefore  
the deficit of events $\sim 0.67$ corresponds  to large 
or nearly maximal mixing.\\ 





\subsection{MINOS}

MINOS  (Main Injector Neutrino Oscillation Search) 
is the long baseline experiment ``from Fermilab to SOUDAN''.
NuMI (Main Injector) beam consists, mainly,  of $\nu_{\mu}$'s 
with energies (1 - 30) GeV and the flux-maximum  at $\sim 3$ GeV.
There are two detectors - steel-scintillator tracking calorimeters.
The near detector is at the distance 1 km from the injector with mass
1 kton and the far detector (SOUDAN mine)  has  the baseline 735 km and 
mass 5.4 kt. The first result corresponds to exposure
$1.27 \cdot 10^{20}$ protons on target. 
215 neutrino events have been observed below 30 GeV, 
whereas $336.0 \pm 14.4$ events were predicted on the basis
of measurements in the near detector \cite{minos-osc}.

\begin{figure}[htb]
\includegraphics[width=80mm,angle=000]{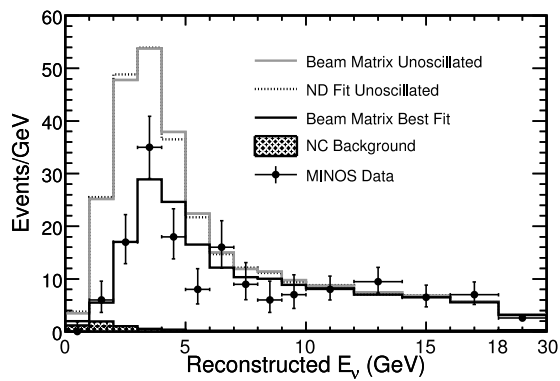}
\caption{The energy spectrum of events in the MINOS experiment, from
\cite{minos-osc}. Shown also are  the expected spectrum without oscillations and 
the best fit to experimental result.}
\label{minos-sp}
\end{figure}

Evidence of oscillations consists of (i) deficit of the
detected number of events - disappearance of the  $\nu_{\mu}$-flux,
and (ii)   distortion of energy spectrum  
fig. \ref{minos-sp}.
The relative suppression increases with decrease of energy
(although there is large spread of points), and
the strongest suppression is in the bins (1 - 2) GeV. 

The dominant effect is the non-averaged 
vacuum $\nu_\mu - \nu_\tau$ 
oscillations. The m atter effect is negligible.
Taking $E \sim 1.5$ GeV as the energy of
the first oscillation dip (minimum) we find  
$\Delta m^2 \sim (2.5 - 3.0) \cdot 10^{-3}$ eV$^2$.
Suppression in this bin is  consistent with maximal mixing.
Using the total deficit of events one can evaluate
the mixing angle more precisely.

Essentially we observe the high energy part of the first oscillation deep 
starting from minimum which  is consistent with maximal suppression. 
This is enough to make rather precise determination of $\Delta m^2$. 
Detailed statistical fit gives $\Delta m^2_{23} = 2.74 ^{+0.44}_{-0.26}
\times 10^{-3}$ eV$^2$ (68 \% C.L.) and
$\sin^2 2 \theta_{23} > 0.87$ (68 \% C.L.) with the best fit value
$\sin^2 2 \theta_{23} = 1$.
This is in a very good agreement with the atmospheric neutrino
and K2K results.

{\it Comment.}
Simple relations we have presented in sect. 4.1 - 4.5  
allow us to understand where sensitivity to different 
parameters comes from.  
These relations are embedded in precise statistical analysis.  
They allow us to control the outcome of this analysis, understand 
uncertainties  and give confidence in the results of more sophisticated 
analysis. 

They show robustness of the results and their interpretation.

\subsection{1-3 mixing: effects and bounds}


The direct bounds on 1-3 mixing are  obtained in the 
CHOOZ experiment~\cite{CHOOZ}. This is the experiment with a single reactor,   
single detector and the baseline about 1 km. 
The expected effect is the vacuum non-average  oscillations with 
survival probability given by the standard oscillation formula:   
\be
P = 1 - \sin^2 2\theta_{13} \sin^2 \frac{\phi}{2}. 
\label{oscpr}
\ee
The  baseline is comparable with the half-oscillation length: 
For the best fit  value of $\Delta m^2$ from the atmospheric neutrino studies  
and $E \sim 2$ MeV  the oscillation length  equals $\sim 2$ km.

The signature of the oscillations consists of distortion
of the energy spectrum described by (\ref{oscpr}). 
No distortion has been found within the  error bars which put the limit  
\be
\sin^2 \theta_{13} \leq  0.04,  ~~(90 \% ~{\rm C.L.})  
\ee
for $\Delta m^2_{31} = 2.6 \cdot 10^{-3}~ {\rm eV}^2$.   
In the atmospheric neutrinos 
the non-zero 1-3 mixing will lead to oscillations of the 
electron neutrinos. One of the effects would be 
$\nu_{\mu}  \leftrightarrow \nu_e$ oscillations in the 
matter of the Earth. 
The resonance enhancement of oscillations in neutrino or antineutrino 
channels should be observable depending on the type of mass 
hierarchy. 
That can  produce an excess of the $e$-like events  
mostly in multi-GeV range where the mixing can be  matter enhanced. 
No substantial effect is found.  
Notice that in the analysis \cite{bari} -  the best fit  value 
$\sin \theta_{13}$ 
is non-zero due to some distortion of the zenith angle dependence. 
in the multi-GeV range.  

In  solar neutrinos, the non-zero 1-3 mixing leads 
to the averaged vacuum oscillations with small oscillation depth. 
The effect is reduced to change of the overall normalization of the 
flux. The combined analysis of all solar neutrino data leads to  zero best-fit value 
of 1-3 mixing. The CC/NC ratio  at SNO and Gallium results 
(which depend on the astrophysical uncertainties less)
give $\sin^2\theta_{13} = 0.017 \pm 0.026$ (that indicates a level of sensitivity of 
existing observations). 
%

In contrast, $\theta_{13}$  can produce
leading effects for supernova electron (anti) neutrinos. 

\subsection{Degeneracy of oscillation parameters and global fits}

In the previous sections we have analyzed various data in the 
$2\nu$ context. Essentially the $3\nu$ system splits in to two sectors: 
``solar'' sector probed by the solar neutrino experiments and KamLAND, and 
the ``atmospheric'' sector probed by the atmospheric neutrino experiments 
K2K and MINOS. This is justified if the 1-3 mixing is zero or small and if  
in the atmospheric sector studies the effect of 1-2 sector can be 
neglected. That could happen, e.g., because in the specific experiments 
the baselines are small or the energies 
are large, so that  oscillation effects 
due to 1-2 mixing and 1-2 split have no time (space) to develop. 

In the next phase of  studies when  
sub-leading effects, {\it e.g.}, induced by $\sin \theta_{13}$, 
become important  the splitting  of $3\nu$ problem into two sectors 
is not possible.  
At this sub-leading level the problem of determination 
of the neutrino parameters becomes much more complicated.  
\begin{table}[b]
\caption{Experiments and relevant oscillation parameters.}
\label{tab2}
\tabcolsep=0.1cm
\begin{tabular*}{\columnwidth}{@{\extracolsep{\fill}}@{}lll@{}}
\hline
Experiments & Parameters of    & Parameters of \\
\,          & leading effects   & sub-leading effects \\
\hline
Solar neutrinos,   & $\Delta m^2_{12}$,  $\theta_{12}$ & $\theta_{13}$ \\
KamLAND                    & \,                                &   \,   \\
Atmospheric neutrinos &  $\Delta m^2_{23}$,  $\theta_{23}$ &  
$\Delta m^2_{12}$,  $\theta_{12}$,  $\theta_{13}$, $\delta$\\
K2K   & $\Delta m^2_{23}$,  $\theta_{23}$ & $\theta_{13}$\\
CHOOZ & $\Delta m^2_{23}$,  $\theta_{13}$ & strongly suppressed \\
MINOS & $\Delta m^2_{23}$,  $\theta_{23}$ & $\theta_{13}$\\
\hline
\end{tabular*}
\end{table}

In the table \ref{tab2}  we indicate relevant parameters for 
different studies. 
The same observables depend on several parameters,  so that the 
problem of degeneracy of the parameters appears. 
%
In such a situation one needs to perform the global fit 
of all available data.  
The advantages are 
(1) no information is lost; 
(2) dependence of different observables on the same 
parameters is taken into account;  
(3) correlation of parameters and their degeneracy 
is adequately treated. 

There are however some disadvantages.   
In particular, for some parameters the global fit may not be the most 
sensitive  method, and   certain subset of the data 
can restrict a given parameter much better  
({\it e.g.}, $\Delta m^2_{23}$ in atmospheric neutrinos).    

\begin{figure}[t]
\includegraphics[width=70mm,angle=000]{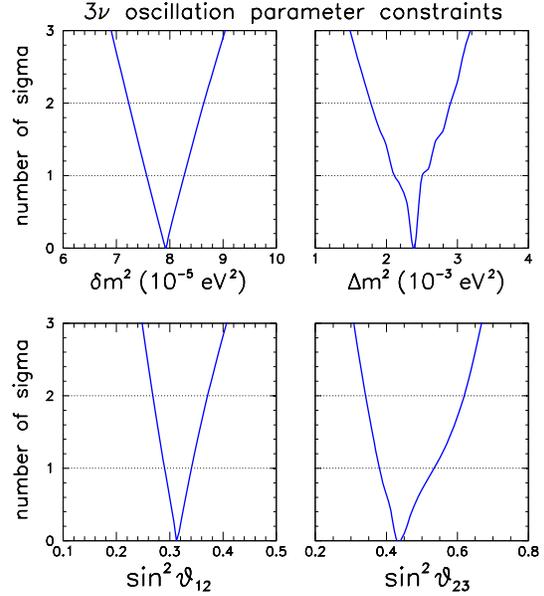}
\caption{The results of global $3\nu$ analysis for 1-2 and 2-3
mass splits and mixings; from \cite{bari}.}
\label{glob}
\end{figure}

In fig. \ref{glob} we show the results of the global fit 
of oscillation data performed in~\cite{bari} before MINOS results. 
MINOS shifts the  allowed region and the best fit point of $\Delta m^2_{23}$ to larger values.  
With earlier MINOS result  
$\Delta m^2_{23} = 2.6 \cdot 10^{-3}~~ {\rm eV}^2$ is found in ~\cite{valencia}
as the best fit value.

Results of global fits of the other groups (see \cite{sv,valencia}) agree very  well.  
Different types of experiments confirm each other: 
KamLAND confirms solar neutrino results, 
K2K - the atmospheric neutrino results {\it etc.}. 
Furthermore, unique interpretation of whole bulk of the data 
in terms of vacuum masses and mixings provides with the overall 
confirmation of the picture
So, the determination of the parameters is rather robust,  
and it is rather non-plausible that future measurements will lead to 
significant change. 

The most probable values of parameters equal 
\begin{eqnarray}
\Delta m^2_{12} = (7.9 - 8.0) \cdot 10^{-5}~~ {\rm eV}^2,\\ 
\sin^2 \theta_{12} = 0.310 - 0.315,~~~~~~~~~~~~~\\
\Delta m^2_{23} = 
(2.5 - 2.6)\cdot 10^{-3}~~ {\rm eV}^2\\ 
\sin^2 \theta_{23} = 0.44 - 0.50.~~~~~~~~~~~~~~~
\end{eqnarray}
Slightly smaller value of 1-2 mixing, 
$\sin^2 \theta_{12} = 0.30$,  has been found in\cite{valencia}. 

The parameter which describes the deviation of the 
23 mixing from maximal equals
\be
D_{23} \equiv 0.5 - \sin^2 \theta_{23} = 0.03 - 0.06. 
\ee
For 1-3 mixing we have 
\be
\sin^2 \theta_{23} = 0.00 - 0.01,  ~~~ (1\sigma =  0.011 - 0.013).  
\ee

The ratio of mass squared differences important for theoretical 
implications equals
\be 
r_{\Delta} \equiv \frac{\Delta m^2_{12}}{\Delta m^2_{23}} = 0.031 - 0.033.  
\label{massre}
\ee


\section{Neutrino mass and flavor spectrum}

\subsection{Spectrum} 
\begin{figure}[htb]
\includegraphics[width=75mm,angle=000]{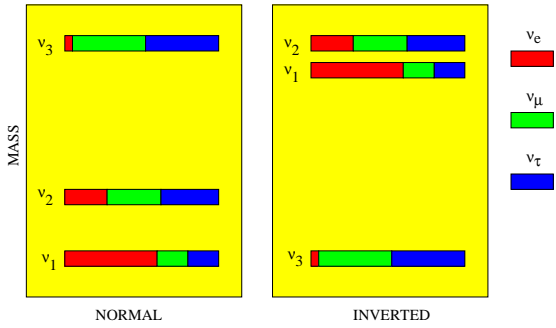}
\caption{Neutrino mass and flavor spectra for the normal (left) and inverted (right) 
mass hierarchies. The distribution of flavors 
(colored parts of boxes) in the mass eigenstates 
corresponds to the best-fit values of mixing parameters  
and  $\sin^2 \theta_{13} = 0.05$. 
}
\label{sp}
\end{figure}

Information obtained 
from the oscillation experiments 
allows us  to make significant progress 
in reconstruction of the neutrino mass and flavor  
spectrum  (Fig.~\ref{sp}).

The unknowns are: 

(i) admixture of $\nu_e$ in $\nu_3$,   $U_{e3}$; 

(ii) type of mass spectrum:
hierarchical,  non-hierarchical with certain ordering, degenerate, 
which is related to the value of the absolute mass scale, $m_1$; 

(iii) type of mass hierarchy (ordering): normal, inverted 
(partially degenerate);  

(iv) CP-violation phase $\delta$.\\

Information described in the previous sections  
can be summarized in the following way. 

1. The observed ratio  of the mass squared differences (\ref{massre}) 
implies that there is no strong hierarchy of neutrino masses: 
\begin{equation}
\frac{m_2}{m_3} >  \sqrt{\frac{\Delta m^2_{12}}{\Delta m^2_{23}}} = 
0.18 \pm 0.02. 
\label{eq:hie1}
\end{equation}
For charge leptons the corresponding ratio is 0.06, and even stronger 
hierarchies are observed in the quark sector.

2. There is the bi-large or large-maximal mixing 
between the neighboring families
(1 - 2) and (2 - 3). Still  rather significant deviation of the 2-3 mixing 
from the maximal one is possible. 

3. Mixing between remote (1-3) families is weak.  

To a good approximation the mixing matrix has the so-called 
tri-bimaximal form \cite{tbm}:    
\be
U_{tbm} \equiv U_{23}^m U_{12}(\theta_{12}) =
\frac{1}{\sqrt{6}}
\left(\begin{array}{ccc}
2 & \sqrt{2} & 0\\
-1 & \sqrt{2} & \sqrt{3}\\
 1 & - \sqrt{2} & \sqrt{3}
\end{array}
\right),
\label{tbimax}
\ee
where $U_{23}^m$ is the maximal ($\pi/4$) rotation in the 
2-3 plane and  $\sin^2 \theta_{12} = 1/3$.
Alternatively, the mixing  can be expressed in terms of  
the quark-lepton complementarity (QLC) relations \cite{qlc}: 
\be
``{\rm lepton~ mixing} =  {\rm bi-maximal~mixing} - {\rm CKM}''.
\ee
Possible realizations are 
\bea
U_{PMNS} =  U_{bm} U_{CKM}^{\dagger}, {\rm or} 
\nonumber \\
U_{PMNS} = U_{CKM} U_{bm} ~~~
\label{order}
\eea
where $U_{CKM}$ is the quark mixing matrix and 
$U_{bm}$ is the bi-maximal mixing matrix: 
\be
U_{bm} \equiv U_{23}^m U_{12}^m =
\frac{1}{2}
\left(\begin{array}{ccc}
\sqrt{2} & \sqrt{2} & 0\\
-1 & 1 & \sqrt{2}\\
1 & - 1 & \sqrt{2}
\end{array}
\right).
\label{bimax}
\ee
Both the tri-bimaximal mixing and the QLC-mixing 
agree with the experimental data within $1\sigma$ \cite{smsm}. 

\subsection{Absolute scale of neutrino mass}

{\it Direct kinematic methods.} Measurements of the Curie plot 
of the $^3H$  decay at  the end point - give 
$m_e < 2.05~ {\rm eV}~~(95 \%)$ (Troitsk)
after ``anomaly''  subtraction \cite{troitsk},  
and $m_e < 2.3~ {\rm eV}~~(95 \%)$,  (Mainz) \cite{mainz}. 
Future KATRIN experiment \cite{katrin} aims at one order of magnitude better  
upper bound:  
$m_e < 0.2~ {\rm eV}~~(90 \%)$. 
The discovery potential is estimated so that  
the positive result $m_e = 0.35$ eV can be established at $5\sigma$ 
(statistical) level. \\

From oscillation experiments we get the lower bound on  
mass of the heaviest neutrino: 
\be 
m_h > \sqrt{\Delta m^2_{atm}} = 0.04~ {\rm eV}~~~ (95\%). 
\ee
In the case of normal mass hierarchy 
$m_h = m_3$ and in the inverted hierarchy case 
$m_h = m_1 \approx m_2$.\\

   
{\it Neutrinoless double beta decay. }
The rate neutrinoless double beta decay 
is determined by the effective Majorana mass of  electron neutrino 
\begin{equation}
m_{ee} =  \left|\sum_k U_{ek}^2 m_k e^{i\phi(k)} \right|, 
\label{eq:it}
\end{equation}
$\Gamma \propto m_{ee}^2$. 
Here $\phi(k)$ is the phase of the $k$ eigenvalue.

The best present bound on $m_{ee}$ 
is given by the Heidelberg-Moscow experiment: 
$m_{ee} < 0.35 - 0.50$ eV~\cite{HM-neg}. Part 
of collaboration claims an evidence of a positive 
signal~\cite{HM-pos,hm}.  
The Heidelberg-Moscow collaboration searched for the   
mode of the decay 
\be
^{76}Ge \rightarrow ^{76}Se +  e^- +  e^-
\ee
with the end point $Q_{ee} = 2039$ keV. 
The total statistics collected from  
5 enriched $Ge-$detectors is 71.7 kg yr. 
The peak at the end point of the total energy-spectrum of two electrons has been found 
and interpreted in \cite{hm} as 
due to the neutrinoless double beta decay. 
Number of events in the peak 
gives the half-lifetime 
\be 
T_{1/2} = 1.19 \cdot 10^{25} ~{\rm y}, ~~~~~3\sigma~{\rm range}:  
(0.69 - 4.18) \cdot 10^{25} ~~{\rm y}. 
\ee
The significance of the peak depends on  model of background and 
quoted by the authors as  $4.2\sigma$. 
There is a number of arguments {\it pro and contra} of such 
interpretation.

If the exchange of light Majorana neutrino is the dominant 
mechanism of the  decay, the measured life-time corresponds to 
the effective mass of the Majorana neutrino  
\be 
m_{ee} = 0.44~ {\rm eV}, ~~~~
3\sigma~{\rm range}:   (0.24 - 0.58)~{\rm eV}. 
\ee
The H-M positive result would correspond 
to strongly degenerate neutrino mass spectrum.  
That, in turn,  implies new symmetry in the leptonic sector. 

Other groups  do not see signal of the  $\beta\beta_{0\nu}$ 
decay though their sensitivity is somehow lower \cite{igex,nemo3,cuoricino}. 
New experiment with $^{76}Ge$, GERDA \cite{gerda}, 
will be able to confirm the H-M claim in the first phase,  
and in the case of negative result, strongly restrict it  in future measurements.\\ 


{\it Cosmological bound.} Analysis of the cosmological data that includes  
CMB observations,  SDSS of galaxies, Lyman alpha forest observations and weak lensing 
lead to the upper bound  on the  sum of neutrino
masses 
\be  
\sum_{i = 1}^3 m_i  < 0.42~ {\rm eV}~~ (95\%~ {\rm C.L.})
\ee 
\cite{cos} (see also \cite{goobar}) which corresponds to $m_0 < 0.13$ eV
in the case of a degenerate spectrum.  An even stronger bound,
$ \sum_{i = 1}^3 m_i  < 0.17~ {\rm eV}$ ($95\%$ C.L.) \cite{Seljak06}
was established after publication of WMAP3 results.
This limit disfavors a strongly degenerate mass spectrum.
and the positive claim of observation of neutrinoless double beta decay. 
Combining the cosmological and  oscillation 
bounds,  we conclude that at least one neutrino mass should be in
the interval
\be
m \sim  (0.04 - 0.10) ~ {\rm eV} ~~~(95\% ~~ {\rm C.L.}).
\ee


In future, the weak lensing will allow to perform direct measurements 
of clustering of all  matter  and not just luminous one. This 
will  improve the sensitivity down to 
$\sum_i m_i  \sim 0.03$ eV. 

\subsection{To the new phase of the field}

In what follows we summarize the parameters, 
physics goals and physics reach of the next generation 
(already approved) LBL experiments. 
In each case we give the  baseline, 
$L$, the mean energy  of neutrino $\langle E_{\nu} \rangle$ and  
the goals. All the estimations are given for the 
$90\%$ C.L..\\

1). T2K (``Tokai to Kamioka''): JPARK $\rightarrow$ SuperKamiokande 
\cite{T2K}. 
This is the accelerator off-axis  experiment on searches for 
$\nu_{\mu} \rightarrow \nu_{\mu}$ and $\nu_{\mu} \rightarrow \nu_e$ 
oscillations; parameters of the experiment:  $L = 295$ km, $\langle 
E_{\nu} \rangle = 0.7$ GeV. 
The goal is to reach sensitivity to the $\nu_e-$appearance which will allow to 
put the bound  
$\sin^2 \theta_{13} < 0.005$ (or discover the 1-3 mixing 
if the angle is larger), to measure 2-3 mass split and mixing with 
accuracies $\delta(\Delta m^2_{23}) \sim 0.1$ meV, 
and $\delta(\sin^2 2\theta_{23}) = 0.01$ near the maximal mixing. 
The latter corresponds to $\delta(\sin^2 \theta_{23}) =  D_{23} = 0.05$.
If 1-3 mixing is near the present bound the hope is to get some information 
about the mass hierarchy.  The measurements will start in 2009. \\
 
2). NO$\nu$A: Fermilab $\rightarrow$ Ash River \cite{nova}.  
This is also the accelerator off-axis experiment on 
$\nu_{\mu} \rightarrow \nu_{\mu}$ and $\nu_{\mu} \rightarrow \nu_e$ 
oscillation searches. 
Parameters:  $L = 810$ km, $\langle E_{\nu} \rangle = 2.2$ GeV. 
The objectives include  the  bound  on 1-3 mixing $\sin^2 \theta_{13} < 0.006$,  
precise measurement of $\Delta m^2_{23}$, and possibly,   determination of 
the mass hierarchy. Start: 2008 - 2009.\\

3). Double CHOOZ reactor experiment \cite{DC} will search for $\bar{\nu}_e 
\rightarrow \bar{\nu}_e$ oscillation disappearance. Two detectors setup 
will be employed.  
Parameters:  $L = 1.05$ km (far detector), $\langle E_{\nu} \rangle = 0.004$ GeV,  
$L/E = 250$ km/GeV; 
the goal is to put  the bound   $\sin^2 \theta_{13} < 0.005 - 0.008$. 
Start: 2008; results: 2011.\\

4). Daya Bay \cite{dayabay}  reactor experiment will  
search for  $\bar{\nu}_e \rightarrow \bar{\nu}_e$ oscillation disappearance 
with multi-detector setup: 
Two near detectors and one far detector with the baseline 1600 - 1900 m 
from reactor cores are proposed. 
The goal is to reach sensitivity   $\sin^2 \theta_{13} < 0.0025$  or better.
Start: 2010.\\

To large extend  results from these experiments 
will determine further (experimental) developments.


\subsection{Expecting the supernova neutrino burst}

Detection of the Galactic supernova can substantially 
contribute to determination of the neutrino parameters 
and reconstruction of the neutrino mass spectrum.
In particular this study will contribute to 
determination of the 1-3 mixing and type of the neutrino mass 
hierarchy.   

In supernovas one expects new elements of the flavor conversion  dynamics. 
Whole $3\nu$ level crossing scheme can be probed  
and  the effects of both MSW resonances (due to $\Delta m^2_{12}$ and $\Delta m^2_{13}$)  
should show up. 
Various effects associated to the 1-3 mixing 
can be realized, depending on value of $\theta_{13}$. 
The SN neutrinos are sensitive to 
$\sin^2\theta_{13}$ as small as $10^{-5}$. 
Studies of the SN neutrinos   will also give an information 
the type of  mass hierarchy~\cite{DS,nun,barg,snsato}.  
The small mixing MSW conversion 
can be realized due to the 1-3 mixing and the ``atmospheric''
mass split $\Delta m^2_{13}$. 
The non-oscillatory adiabatic conversion 
is expected for $\sin^2\theta_{13} > 10^{-3}$.  
Adiabaticity violation 
occurs if the 1-3 mixing is small $\sin^2\theta_{13} < 10^{-3}$.

Collective flavor transformation effects due to the 
neutrino-neutrino scattering (flavor exchange phenomenon)
can be important in the central parts (outcide neutrino spheres) of the 
collapsing stars \cite{collect}.

Another possible interesting effect is related to shock wave propagation. 
The shock wave can reach the region
of the neutrino conversion, $\rho \sim 10^{4}$ g/cc,
after $t_s = (3 - 5)$ s from the bounce 
(beginning of the $\nu-$ burst)~\cite{SF}.
Changing suddenly the density profile and therefore breaking the  
adiabaticity, 
the shock wave front influences the conversion in
the resonance characterized by  $\Delta m^2_{13}$ and $\sin^2 \theta_{13}$,
if $\sin^2 \theta_{13}  > 10^{-4}$.

Monitoring the shock wave with neutrinos 
can shed some light on the mechanism 
of explosion~\cite{Tak,DS,lisis,munich}. 


\subsection{LSND result and new neutrinos}

LSND  (Large Scintillator Neutrino Detector) collaboration 
studied interactions of neutrinos from
Los Alamos Meson Physics Facility produced in 
the decay chain:  $\pi^+ \rightarrow \mu^+ + \nu_e$,
$\mu^+ \rightarrow e^+ + \nu_e + \bar{\nu}_{\mu}$.
The excess of the $(e^+ + n)$ events has been observed in the detector
which could be due to inverse beta decay: $\bar{\nu}_e + p
\rightarrow e^+ + n$ \cite{LSND}. In turn,  $\bar{\nu}_e$ could appear
due to oscillations  $\bar{\nu}_{\mu} \rightarrow \bar{\nu}_e$ in  the original
$\bar{\nu}_{\mu}$ beam.

Interpretation of the excess 
in terms of the $\bar{\nu}_{\mu} - \bar{\nu}_e$ oscillations 
would correspond to the transition probability
\be
P = (2.64 \pm 0.76 \pm 0.45) \cdot 10^{-3}. 
\ee
The allowed region is restricted from below by 
$\Delta m^2 > 0.2$ eV$^2$.  

This result is clearly beyond the ``standard'' $3\nu$ picture. 
It implies new sector and new symmetries of the theory. 

The situation with this ultimate neutrino anomaly \cite{LSND} 
is really dramatic: all suggested 
physical (not related to the LSND methods) 
solutions are strongly or very strongly disfavored now. 
At the same time, being confirmed, the oscillation interpretation of  
the LSND result  may change our understanding the  neutrino 
(and in general fermion) masses. 

Even very exotic possibilities are disfavored. 
An analysis performed by the KARMEN collaboration \cite{KARMEN} 
has further disfavored a scenario \cite{BaP} 
in which the $\bar{\nu}_e$ appearance  is explained by the 
anomalous muon decay  $\mu^+ \rightarrow \bar{\nu}_e \bar{\nu}_{i} e^+$  
$(i = e , \mu, \tau)$. 

The CPT-violation scheme\cite{BLyk}  with different mass spectra 
of neutrinos and antineutrinos 
is disfavored by the atmospheric neutrino data~\cite{stru}. 
No compatibility of LSND and 
``all but LSND" data have been found below $3\sigma$~\cite{coCPT}.  

The main problem of the (3 + 1) scheme with 
 $\Delta m^2 \sim 1$ eV$^2$  is that the predicted LSND signal,
which is consistent with the results  of other short base-line experiments
(BUGEY, CHOOZ, CDHS, CCFR, KARMEN) 
as well as the atmospheric neutrino data,  is too small:
the $\bar{\nu}_{\mu} \rightarrow \bar{\nu}_e$ probability 
is about $3\sigma$ below the  LSND measurement.

Introduction of the second sterile  neutrino 
with $\Delta m^2 >  8$ eV$^2$ may help \cite{PS31}. 
It has been  shown~\cite{sorel}  that a new neutrino with 
$\Delta m^2 \sim 22$ eV$^2$ and mixings $U_{e5} = 0.06$, $U_{\mu5} = 0.24$
can enhance the predicted LSND signal by (60\,--\,70)\%.
The (3 + 2) scheme has, however, problems with cosmology and astrophysics. 
The combination of the two described solutions, namely the $3 + 1$ 
scheme with CPT-violation  has been considered~\cite{barger3}. 
Some recent proposals including the  mass varying neutrinos MaVaN~ 
\cite{mavan}  and decay of heavy sterile neutrinos~\cite{lsndste} also have certain 
problems.

MiniBooNE ~\cite{miniboone} is expected to clarify substantially 
interpretation of the LSND result. 
The MiniBooNE  searches for $\nu_e$ appearance in the 12 m diameter 
tank filled in by the 450 t 
of mineral oil scintillator and covered by 1280 PMT.  
The flux of muon neutrinos with the average energy 
$\langle E_{\nu} \rangle \approx 800$ MeV
is formed in $\pi$ decays (50m decay pipe) which are in turn produced by 
8 GeV protons from the Fermilab Booster. The 541 m baseline is 
about half of the oscillation length for $\Delta m^2 \sim 2$ eV$^2$. 
In 2006 the experint operates in the antineutrino channel 
$\bar{\nu}_\mu \rightarrow \bar{\nu}_e$

\begin{figure}[t]
\centering
\includegraphics[width=60mm,angle=000]{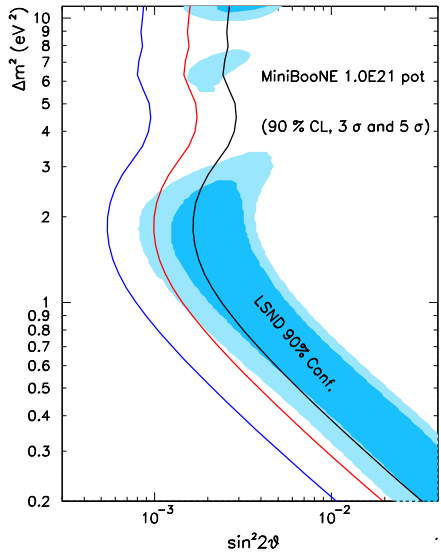}
\caption{The region of oscillation parameters selected by LSND result
versus sensitivity of the MiniBooNE experiment; from~\cite{miniboone}.}
\label{mboone}
\end{figure}

Of course, confirmation of the 
LSND result (in terms of oscillations) would be the most decisive 
though the problem with background should be scrutinized.  
The negative result still may leave an ambiguous  situation.  
In fig. \ref{mboone} the sensitivity limits and discovery potential 
of MiniBooNE  are shown.

\section{Conclusion}

After the first phase of studies  of neutrino mass and mixing 
we have rather consistent picture: 
interpretation of all the results 
(except for LSND) in terms of vacuum mixing
of three massive neutrinos.
Two  main effects (consequences of  mixing) 
are important for the interpretation 
of results at the present level of accuracy: the vacuum oscillations and the adiabatic
conversion in matter (the MSW-effect). 
The oscillations in matter give sub-leading contributions, at $(1-2)\sigma$ level,  to the  solar and 
atmospheric neutrino observables. 

There are unknown yet parameters and their
determination  composes a program of future phenomenological 
and experimental studies.
Next phase of the field, study of sub-leading effects, will be much more involved.

The main theoretical challenge 
is to understand what is behind the observed
pattern of neutrino masses and  mixing (as well as masses
and mixings of other fermions).  
What is the underlying physics?
Clearly there is a strong difference of the quark 
and lepton mixing patterns.   
The data hint the tri-bimaximal scheme of mixing 
with  possible implications of new ``neutrino 
symmetries'',  
or alternatively to  the quark-lepton complementarity that hints certain quark-lepton symmetry and 
unification.  
Are the tri-bimaximal or QLC relations real (follow from certain principles) 
or simply accidental?

It may happen that something important 
in principles and context is still missed.
The key question is how far we can go in this understanding using 
our usual notions of the field theory (or the effective field theory) and
in terms of symmetries, various mechanisms of
symmetry breaking, {\it etc.}?
The hope is that neutrinos will  uncover something simple 
and illuminating before we will be lost in the 
string landscape. \\



%
\end{document}